\documentclass[
  aps,
  prd,
  reprint,        
  amsmath,amssymb,
  nofootinbib,
  superscriptaddress
]{revtex4-2}

\usepackage{graphicx}
\usepackage{subcaption}
\usepackage{bm}
\usepackage{booktabs}
\usepackage{siunitx}
\usepackage{hyperref}      

\hypersetup{colorlinks=true, linkcolor=blue, citecolor=blue, urlcolor=blue}
\graphicspath{{figures/}}

\newcommand{\Mhm}{M_{\rm hm}}
\newcommand{\mchi}{m_\chi}
\newcommand{\deff}{d_{\rm eff}}
\newcommand{\xmin}{\bm{x}_{\rm min}}
\newcommand{\Amin}{\mathcal{A}_{\rm min}}
\newcommand{\msub}{m_{\rm sub}}
\newcommand{\dps}{\delta\psi}
\newcommand{\RF}{R_F}
\newcommand{\Rnear}{R_{\rm near}}
\newcommand{\msubmin}{\msub^{\rm min}}

\begin{document}

\title{Wave-Optics Imprints of Warm Dark Matter Subhalos with Prompt Cusps on Strongly Lensed Gravitational Waves}

\author{Bhashin A. Thakore}
\email{bhashinashish.thakore@unito.it}
\affiliation{Dipartimento di Fisica, Universit\`a degli Studi di Torino, Via P.\ Giuria 1, 10125 Torino, Italy}
\affiliation{INFN -- Istituto Nazionale di Fisica Nucleare, Sezione di Torino, Via P.\ Giuria 1, 10125 Torino, Italy}
\affiliation{GRAPPA (GRavitation AstroParticle Physics Amsterdam), University of Amsterdam, Science Park 904, 1098 XH Amsterdam, The Netherlands}

\author{Shin'ichiro Ando}
\email{s.ando@uva.nl}
\affiliation{GRAPPA (GRavitation AstroParticle Physics Amsterdam), University of Amsterdam, Science Park 904, 1098 XH Amsterdam, The Netherlands}
\affiliation{Kavli Institute for the Physics and Mathematics of the Universe, University of Tokyo, Chiba 277-8583, Japan}
\begin{abstract}
Dark matter halos are expected to form with a prompt $\rho \propto r^{-3/2}$ density cusp at their centres, and in warm dark matter (WDM) cosmologies these cusps can dominate the inner structure of the low-mass subhalos that survive free-streaming suppression. We investigate whether that inner structure is visible to gravitational waves lensed in the wave-optics (WO) regime. Extending the strong lensing WO diffraction-integral framework to WDM, we generate subhalo populations with the {\sc sashimi-w} semi-analytic model, assign each subhalo a prompt cusp through a cusp-halo relation, and compute the frequency-dependent amplification factor $F(f)$ across the LISA band over 500 independent realisations for $m_\chi \in \{6, 10, 20, 40, 80, 100\}$~keV. The signal is governed by the overlap between the WDM-suppressed subhalo mass function and the $m_{\rm sub} \sim 10^{4}$ to $10^{7}\,M_\odot$ range to which the LISA band is sensitive. Modulations are negligible for $m_\chi \lesssim 10$~keV, reach the percent level by 40~keV, and then saturate to a CDM-like plateau, so the observable reliably probes $m_\chi$ only over the intermediate range $\sim 20$ to 40~keV. Repeating the 40~keV ensemble on identical subhalo catalogs with the cusp amplitude set to zero, we find that prompt cusps enhance the amplitude modulation only mildly. Where cusps make up a large fraction of the subhalo mass, free-streaming has already depleted the signal-carrying mass range, and where that range is populated, the cusp lies well inside the Fresnel radius, to which the diffraction integral responds through projected enclosed mass rather than central density. A percent-level detection in a strongly lensed massive black-hole binary would therefore establish the presence of substructure on Fresnel scales without diagnosing its inner profile.
\end{abstract}

\maketitle

\section{Introduction}
\label{sec:intro}

The presence of dark matter (DM) substructure is an  outcome of significant importance to hierarchical structure formation, and provides a sensitive observational window into the particle-physics properties of DM on sub-galactic scales \cite{Springel2008, Bullock2017}. Within the cold dark matter (CDM) framework, halos surrounding individual galaxies are predicted to contain a rich population of bound subhalos spanning a wide range of masses, extending down to arbitrarily small scales. In WDM scenarios, by contrast, free-streaming of the light, thermal-relic particle washes out primordial density perturbations below a characteristic length scale, which in turn suppresses both the number and central density of halos with masses below the half-mode mass
$\Mhm$ \cite{Lovell2014, Bose2016}. Detecting the extent of this suppression, therefore, provides a direct handle on the mass of the DM particle.

Several probes of small-scale structure have studied the WDM free-streaming scale, including galaxy-scale strong gravitational lensing \cite{Vegetti2024,Gilman2020} and the abundance and internal dynamics of Milky Way satellite galaxies \cite{Simon2019, nadler2021constraints}. These methods are powerful but subject to astrophysical systematics and uncertainties. A probe that responds directly to the gravitational potential of dark substructure, with minimal baryonic contamination, would provide a valuable independent cross-check.

DM halos can be expected to form with a prompt $\rho \propto r^{-3/2}$ density cusp at their centres, produced by the smooth gravitational collapse of the initial density peaks from which they formed. Inner cusps close to this slope, steeper than the NFW form, have been found for more than a decade in high-resolution simulations of the first halos near the free-streaming scale \cite{Ishiyama2010, Anderhalden2013, Ishiyama2014, Angulo2017, Delos2018, delos2019predicting, Colombi2021}, including subhalos \cite{IshiyamaAndo2020}, with further work addressing how such cusps form and how they evolve toward shallower NFW-like profiles as halos grow \cite{Ogiya2016, Angulo2017, OgiyaHahn2018, Colombi2021}. In WDM cosmologies these cusps are of particular interest due to the formation of the surviving low-mass subhalos near the free-streaming scale. These prompt cusps constitute a substantial fraction of their inner mass and partially counteract the WDM suppression of central densities \cite{Delos2025_WDM}. The amplitude of the cusp is tied to the formation history of its host through the cusp-halo relation of Ref.~\cite{delos2025cusp}, which we adopt to assign cusps to our subhalo population.

Gravitational lensing of gravitational waves (GWs) offers a qualitatively new window for further insight into DM substructure. Since the GW wavelength is macroscopic, lensing effects from a low-mass perturber manifest in the wave-optics (WO) regime, in which the phase coherence of the wave produces frequency-dependent interference whenever the lensing time delay is comparable to the wave period \cite{Takahashi2003}. Considerable study has gone into WO signatures from compact objects and low-mass halos across both ground- and space-based detector platforms, including their use as probes of halo density profiles and DM properties \cite{Tambalo2023} and of the solitonic cores of fuzzy DM halos \cite{Singh2025}. However, forecasts for individual lenses in generic line-of-sight configurations suggest that detectable frequency-dependent modulations will arise in only a few events over a given mission lifetime (see Refs.~\cite{Fairbairn2023,Guo2022,brando2025signatures}). A complementary approach targets the cumulative signal of many unresolved lenses along the line of sight, as in lens stochastic diffraction \cite{Zumalacarregui2024} and the ``dark timbre'' of LISA binaries \cite{Urrutia2025}, with recent forecasts for stochastic diffraction by low-mass halos in LISA \cite{Choi2026}. Strongly lensed GWs are different, wherein near a macro-critical curve, the large geometric magnification amplifies the small Fermat-potential perturbations of nearby subhalos, boosting the WO signal to potentially detectable levels. Ref.~\cite{Ando2026} demonstrated that percent-level amplitude and phase modulations arise generically for CDM substructure in this regime, within reach of the Laser Interferometer Space Antenna (LISA) \cite{amaro2017laser}. Additionally, LISA will also offer a well-defined and physically motivated sample of multiply imaged GW events in which WO signatures of dark matter substructure can be systematically searched (Refs.~\cite{Oguri2018,sereno2010strong,gutierrez2025strong}).

In this work we extend the strongly-lensed WO calculation of Ref.~\cite{Ando2026} to WDM. We generate WDM subhalo populations with the SASHIMI-W semi-analytical model \cite{dekker2022warm,hiroshima2018modeling,ludlow2016mass}, assign each subhalo a prompt cusp through the cusp-halo relation of Ref.~\cite{delos2025cusp}, and compute the full frequency-dependent amplification factor $F(f)$ across the LISA band over 500 independent realisations for WDM particle masses at $\mchi \in \{6, 10, 20, 40, 80, 100\}\,\mathrm{keV}$. We show that the WO signal is controlled by the overlap between the WDM-suppressed mass function and the WO-sensitive mass window set by the detector band: it is absent for the lightest WDM candidates, rises to the percent level near $\mchi \sim 40\,\mathrm{keV}$, and saturates to a CDM-like plateau once the half-mode mass falls below the window. We further isolate the role of the prompt cusps using matched subhalo catalogs, finding that they leave the mean magnification unbiased and enhance the amplitude modulation only modestly. We trace this to an opposition between the subhalo masses that carry the WO signal and the WDM models for which prompt cusps are a significant part of those subhalos, which holds across the parameter space rather than only at our fiducial particle mass.

The paper is organized as follows. Section~\ref{sec:model} describes the macrolens configuration, the WDM subhalo population, the prompt-cusp lens potential, and the WO calculation. Section~\ref{sec:results} presents the ensemble statistics of $F(f)$ and their dependence on the minimum subhalo mass, along with an assessment of the prompt-cusp contribution to the amplitude modulation. Section~\ref{sec:discussion} explains why the cusped and cusp-free signals agree so closely, and assesses the detectability of such a signal with LISA. We conclude in Section~\ref{sec:conclusion}.

\section{Model and Formalism}
\label{sec:model}
We consider a strongly lensed GW system consisting of a macroscopic host halo populated with WDM subhalos.  While the macro-image positions and the macrolens Jacobian $\Amin$ are computed using the GLoW framework \cite{villarrubia2025gravitational}, the WO subhalo lens potential is integrated numerically on a customised 2D grid.

\subsection{Macrolens model}
\label{sec:macro}

The macrolens consists of a dark matter halo at redshift $z_L = 0.5$ described by a Navarro-Frenk-White (NFW) profile \cite{NFW1996}, with $M_{200c} = 10^{12}\,M_\odot$ and fiducial concentration $c_{200c} = 10/(1+z_L)$ \cite{correa2015accretion}, following the same macrolens configuration as Ref.~\cite{Ando2026}. A central galaxy is modeled as a singular isothermal sphere (SIS) with velocity dispersion $\sigma_v = 250\,\mathrm{km\,s^{-1}}$. The source is placed at $z_S = 1.5$. 

Subhalos with $\msub > 10^9\,M_\odot$ are included within the macrolens and treated in the geometric-optics (GO) limit. Their contribution perturbs the image positions and local convergence but does not produce frequency-dependent WO distortions in the LISA band.  We focus throughout on the macro minimum image, where any detectable modulation would arise from the local WO subhalo population.
The saddle image, on the other hand, is expected to yield WO modulation at a comparable level~\cite{Ando:2026eam}.

\subsection{Selection radius}
\label{sec:split}
We define the sampling radius for WO subhalos around each image as defined in Ref.~\cite{Ando2026}:
\begin{equation}
    \Rnear = \max\!\left[N_F \RF,\; R_\mu,\; R_{\rm core}\right],
    \label{eq:Rnear}
\end{equation}
where $N_F = N_E = 5$, $\RF = [c\deff/(2\pi f_{\rm min})]^{1/2}$ is the Fresnel radius at $f_{\rm min}$, $R_\mu = [g_{\rm img}\msub/(\pi\epsilon_\mu \Sigma_{\rm cr})]^{1/2}$ is the magnification-perturbation radius requiring a GO fractional perturbation $\epsilon_\mu = 10^{-2}$. The third term is expanded as $ \min(\left[N_{E}r_{s,{\rm sub}}, r_{t,{\rm sub}}\right])$, where $N_E = 5$. This ensures that the sampling radius does not drop below the characteristic scale radius of the truncated NFW subhalo. Subhalo projected positions are drawn from the host NFW radial profile $n_{\rm sub}(r) \propto (r^2 + R_s^2)^{-3/2}$ \cite{Ando2026} within the disc of radius $\Rnear$ centered on the image.

\subsection{WDM subhalo populations}
\label{sec:wdm_subs}

Subhalo populations are generated using SASHIMI-W semi-analytic framework \cite{dekker2022warm,hiroshima2018modeling,ludlow2016mass},\footnote{While the original SASHIMI-W code uses the WMAP power spectrum, the code we currently use has been adjusted for use with the Planck (\cite{aghanim2020planck}) matter power spectrum} which predicts the subhalo mass function and structural evolution within a host halo.  The NFW concentrations follow the concentration-mass relation as implemented within SASHIMI-W, described by Ref.~\cite{ludlow2016mass}. Each catalog entry is characterized by its mass $\msub$, NFW scale radius $r_s$, characteristic density $\rho_s$, and tidal radius $r_t$.

The WDM parameter of note is the half-mode mass $\Mhm$, the scale below which the transfer function suppresses the linear matter power spectrum by a factor of $\geq 2$. We compute $\Mhm$ using the same  thermal-relic free-streaming length (Ref.~\cite{Viel2005}) adopted internally by SASHIMI-W. The representative values are listed in Table~\ref{tab:mhm}. 

The subhalo mass function falls steeply below $\Mhm$, so the number of WO subhalos near the image depends critically on whether $\Mhm$ falls within the WO mass window $[10^2, 10^9]\,M_\odot$. At $\mchi = 6\,\mathrm{keV}$, $\Mhm \sim 10^{7.6}\,M_\odot$ lies near the upper boundary of this window, which admits $\sim 0$ WO subhalos.  At $40$--$100 \,\mathrm{keV}$, $\Mhm$ falls within this window, admitting a slightly larger population.

\subsection{Prompt-cusp lens potential}
\label{sec:cusps}
In addition to the standard truncated NFW profile, each WO subhalo harbours a prompt cusp
arising from the smooth gravitational collapse of its initial progenitor \cite{Delos2018,
delos2019predicting}.  Following \cite{delos2025cusp}, the three-dimensional density profile is
\begin{equation}
    \rho(r) = \frac{\sqrt{y^2 + x}}{x^{3/2}(1+x)^2}\,\rho_s,
    \quad x \equiv \frac{r}{r_s},\quad y \equiv \frac{A}{\rho_s r_s^{3/2}},
    \label{eq:cusp_nfw}
\end{equation}
which transitions from $\rho \propto r^{-3/2}$ (prompt cusp) at $r \ll r_s$ to the standard
NFW form at $r \gg r_s$.  The dimensionless cusp amplitude $y$ (equivalently the physical
amplitude $A$ in $M_\odot\,\mathrm{kpc}^{-3/2}$) is assigned using the cusp-halo relation
of \cite{delos2025cusp}, evaluated at the subhalo's infall mass $m_{\rm infall}$ and accretion
redshift $z_{\rm acc}$ as supplied by SASHIMI-W.  A cusp is assigned only where the
cusp-halo relation predicts one at that infall mass and redshift; below the corresponding
formation threshold, an infall mass of $\approx 0.05\,\Mhm$, we set $A = 0$ and the subhalo
reverts to a plain NFW profile. Because the amplitude is fixed at infall while the lensing uses
the stripped present-day structure, the cusped population extends to present-day masses well below
this threshold.

The cusp amplitude $A$ is set by the halo's formation history through the cusp-halo relation of \cite{delos2025cusp}.  Each prompt cusp arises from the collapse of a local maximum in the primordial density field of characteristic length scale $L = |\delta/\nabla^2\delta|^{1/2}$; at the collapse time $t_{\rm coll}$ the cusp amplitude and enclosed mass are $A = \alpha\,[\bar\rho\,L^{3/2}]_{\rm coll}$ and $m_{\rm cusp} = \beta\,[\bar\rho\,L^3]_{\rm coll}$, with $\alpha \simeq 24$ matched to $N$-body simulations and $\beta \simeq 7.3$ from a theoretical argument \citep{delos2025cusp}.  Newly formed ``young'' cusps follow a scaling relation between $A$ and $m_{\rm cusp}$, which in natural units of the power-spectrum moments $\sigma_j^2 \equiv \int_0^\infty (dk/k)\,\mathcal{P}(k)\,k^{2j}$ reads
\begin{equation}
    \frac{A}{\bar\rho\,\sigma_0^{9/4}\sigma_2^{-3/4}}
    \;\simeq\;
    C\left[\frac{m_{\rm cusp}}{\bar\rho\,(\sigma_0/\sigma_2)^{3/2}}\right]^{\!p},
    \label{eq:Am_relation}
\end{equation}
with $C \simeq 0.8$ and $p \simeq 1.9$, approximately independent of the assumed cosmology
\cite{delos2025cusp}.

To predict $A$ for a halo of present-day mass $M$, one traces its mass accretion history backward to the epoch at which the central cusp first formed.  Reference~\cite{delos2025cusp} shows that halo masses grow in proportion to a universal factor $\chi(\sigma_0) = e^{-\kappa/\sigma_0}$
($\kappa \simeq 4.5$, with $\sigma_0 \propto a$ being the time variable), so that combining the above relations with this growth law yields an equation for the cusp formation epoch $\sigma_0|_{\rm coll}$, solved analytically via the Lambert $W$ function. The resulting $(m_{\rm cusp},\,A)$ prediction reproduces the median cusp coefficient to within $\sim$30 per cent in general, and to $\sim$10 per cent for the WDM-like ($n=-2$) power spectra most relevant here, with an approximately lognormal scatter of $\sigma_{\log_{10}A} \simeq 0.15\,\mathrm{dex}$.  In WDM cosmologies, free-streaming suppresses the small-scale moments $\sigma_j$ and delays the collapse of low-mass peaks. Heavier particles, with their smaller free-streaming scale, produce systematically smaller cusps (lower $A$) throughout the halo population, while for a halo of fixed mass $M$ the central-cusp $A$ decreases with $\mchi$ much more gradually (Ref.~\cite{delos2025cusp}).  Since the relation is calibrated for field halos, we follow the prescription of \cite{delos2025cusp} for subhalos and evaluate $A$ at the mass and redshift the object had before tidal stripping removes the outer envelope while leaving the central cusp largely intact using the infall mass $m_{\rm infall}$ and accretion redshift $z_{\rm acc}$ supplied by SASHIMI-W for each WO subhalo.

The projected lensing potential of the cusp-NFW profile does not admit a closed-form
expression.  We compute it numerically as follows.  For each WO subhalo, the dimensionless
convergence profile is
\begin{equation}
    \kappa(u) = \frac{1}{\Sigma_{\rm cr}} \int_{-\infty}^{\infty}
                \rho\!\left(\sqrt{u^2 + z^2}\right) dz,
    \label{eq:kappa_los}
\end{equation}
where $u$ is the projected radius in the lens plane.  
In the prompt-cusp-dominated regime $r \ll r_s$ (equivalently $x \ll y^2$ in the
notation of Eq.~\ref{eq:cusp_nfw}), the factor $\sqrt{y^2+x}\approx y$ and
$(1+x)^2\approx 1$, so the density simplifies to
\begin{equation}
    \rho(r)\;\xrightarrow{r\ll r_s}\;\frac{y\,\rho_s}{x^{3/2}}
    = \frac{A}{\rho_s r_s^{3/2}}\,\frac{\rho_s}{(r/r_s)^{3/2}}
    = \frac{A}{r^{3/2}}.
    \label{eq:rho_cusp_limit}
\end{equation}
Substituting into Eq.~(\ref{eq:kappa_los}) with $r=\sqrt{u^2+z^2}$, the integrand
near the origin scales as $\rho\sim A(u^2+z^2)^{-3/4}$.  We then substitute $z = ut$, as:
\begin{eqnarray}
    \kappa(u) &\;\approx\;& \frac{2A}{\Sigma_{\rm cr}}
    \int_0^\infty (u^2+z^2)^{-3/4}\,dz 
    \nonumber\\ &=& {}
    \frac{2A}{\Sigma_{\rm cr}}\,u^{-1/2}
    \underbrace{\int_0^\infty (1+t^2)^{-3/4}\,dt}_{\displaystyle\equiv\,\mathcal{C}},
    \label{eq:kappa_scaling}
\end{eqnarray}
where $\mathcal{C} = \tfrac{1}{2}\sqrt{\pi}\,\Gamma(1/4)/\Gamma(3/4)\approx 2.62$ is a
finite numerical constant, confirming $\kappa(u)\propto u^{-1/2}$ as $u\to 0$. 

The deflection angle and projected potential then follow by successive radial integration:
\begin{equation}
    \alpha(u) = \frac{2}{u}\int_0^u \kappa(u')\,u'\,du', \qquad
    \psi(u)   = \int_0^u \alpha(u')\,du',
    \label{eq:alpha_psi}
\end{equation}
evaluated numerically by the trapezoidal rule on the same radial grid. Both integrals are well-defined despite the $u^{-1/2}$ divergence in $\kappa$ as the weight $u'$ in the integrand for $\alpha$ renders it $\propto u^{1/2}$ near the origin, with $\psi$ inheriting a similarly mild cusp.  The resulting 1D profiles $\{\kappa, \alpha, \psi\}(u)$ are stored as interpolation tables and sampled at each grid point via the projected radius $u = |\bm{u} - \bm{u}_{\rm sub}|$, and then added to the Fermat time-delay surface before the numerical integration.

\subsection{Wave-optics calculation}
\label{sec:wocalc}

Introducing physical transverse coordinates $\bm{\xi}$ and $\bm{\eta}$ in the lens and source planes, respectively, we define
dimensionless variables $\bm{x} \equiv \bm{\xi}/\xi_0$ and $\bm{y} \equiv (\bm{\eta}/\xi_0)(D_L/D_S)$, where $\xi_0$ is an arbitrary normalization length. The GW amplification factor is then
\begin{equation}
    F(f) = \frac{w}{2\pi i} \int d^2x\;\exp\bigl[iw\,\phi(\bm{x},\bm{y})\bigr],
    \label{eq:Fw}
\end{equation}
where $w \equiv f/f_0 = 2\pi f\xi_0^2/(c\deff)$ is the dimensionless frequency, $\deff \equiv D_L D_{LS} / [(1+z_L)D_S]$ the effective lensing distance, and $D_L$, $D_S$, $D_{LS}$ the angular-diameter distances to the lens, to the source, and between them. This effective distance is not an independent input, being fixed by the critical surface density through $\deff = c^2/[4\pi G\,\Sigma_{\rm cr}(1+z_L)]$. For $z_L = 0.5$ and $z_S = 1.5$ this gives $\deff \simeq 490\,\mathrm{Mpc}$. We also define the dimensionless Fermat potential as
\begin{equation}
    \phi(\bm{x},\bm{y}) = \tfrac{1}{2}|\bm{x} - \bm{y}|^2 - \psi(\bm{x}),
    \label{eq:fermat}
\end{equation}
 with $\psi$ the total lens potential. The first term in $\phi$ is the geometric path delay, while $-\psi(\bm{x})$ encodes the Shapiro delay from the projected mass distribution. The lensing potential is related to the convergence $\kappa = \Sigma/\Sigma_{\rm cr}$ by
the two-dimensional Poisson equation $\nabla^2_{\bm{x}}\psi = 2\kappa(\bm{x})$, or equivalently
\begin{equation}
    \psi(\bm{x}) = \frac{1}{\pi}\int d^2x'\,\kappa(\bm{x}')\ln|\bm{x}-\bm{x}'|.
    \label{eq:psi_kappa}
\end{equation}
Images form at the stationary points of $\phi$, i.e.\ where $\nabla_{\bm{x}}\phi = 0$, giving the standard lens equation
\begin{equation}
    \bm{y} = \bm{x} - \nabla_{\bm{x}}\psi(\bm{x}).
    \label{eq:lens}
\end{equation}
In the GO limit ($w\to\infty$), Eq.~(\ref{eq:Fw}) is dominated by these stationary points and reduces to
\begin{equation}
    F_{\rm GO}(w,\bm{y}) = \sum_j \sqrt{|\mu_j|}\;
    \exp\!\left[iw\,\phi(\bm{x}_j,\bm{y}) - i\tfrac{\pi}{2}n_j\right],
    \label{eq:Fgeo}
\end{equation}
where $\mu_j = 1/\det \mathcal{A}_j$ is the signed magnification at image $j$,
$\mathcal{A}_j \equiv I - \nabla\nabla\psi|_{\bm{x}_j}$ is the Jacobian of the lens mapping, and for our case consider only the minimum images. For a minimum image, $F_{\rm GO} \equiv 1/\sqrt{\det \mathcal{A}_{\rm min}}$ sets the normalization used in Eq.~(\ref{eq:ratio}) below. 

Defining local coordinates $\bm{u} \equiv \bm{x} - \xmin$ centered on the macro minimum image at $\xmin$, and decomposing the lensing potential as
\begin{equation}
    \psi(\bm{x}) = \psi_{\rm macro}(\bm{x}) + \dps(\bm{x}),
    \label{eq:psi_decomp}
\end{equation}
where $\psi_{\rm macro}$ includes the host NFW halo, SIS galaxy, and massive ($> 10^9\,M_\odot$) subhalos, and $\dps$ contains the WO subhalo contributions, the integral becomes
\begin{equation}
    F(w) = \frac{w}{2\pi i}\int d^2u\;\exp\!\left[iw\!\left(
           \tfrac{1}{2}\bm{u}^T\Amin\bm{u} - \dps(\bm{u})\right)\right],
    \label{eq:Fw_local}
\end{equation}
where $\Amin \equiv I - \nabla\nabla\psi_{\rm macro}\big|_{\xmin}$ is the Jacobian matrix of the macrolens mapping evaluated at the minimum image.  This construction incorporates the dominant macrolens field exactly to quadratic order while treating $\dps$ as the explicit WO perturbation.  

We evaluate Eq.~(\ref{eq:Fw_local}) on a two-dimensional $N = 2048$ grid with a Gaussian apodisation window of frequency-dependent width $\sigma(w) \propto w^{-1/2}$.  To isolate the subhalo contribution we adopt a ratio normalization,
\begin{equation}
    F(f) = F_{\rm GO}\;\frac{F_{\rm full}(f)}{F_0(f)},
    \label{eq:ratio}
\end{equation}
where $F_{\rm GO}$ is the GO macro amplification, $F_{\rm full}$ the full WO result including $\dps$, and $F_0$ the result with $\dps = 0$ on the same grid.  By construction, $F_{\rm full}/F_0 \to 1$ in the smooth limit, so any frequency dependence in the ratio is a direct signature of the subhalo population.

\section{Results}
\label{sec:results}
\subsection{Amplitude Modulation Due to Prompt Cusp-Infused Subhalos}
\label{sec:ensemble}
Similar to the procedure carried out in Ref.~\cite{Ando2026}, Fig.~\ref{fig:ensemble} shows $|F(f)|$ (top), the relative modulation $|F(f)/F(f_{\rm ref})|-1$ normalized at $f_{\rm ref} = 0.1\,\mathrm{Hz}$ (middle), and the phase shift $\arg F(f)$ (bottom) for $m_\chi \in \{6, 10, 20, 40, 80, 100\}\text{ keV}$, each computed over 500 independent subhalo realisations. Shaded bands mark the 68\% and 95\% percentile ranges; the solid black line is the median.

At $m_\chi = 6\,\mathrm{keV}$, both the amplitude modulation and phase shift remain at the $10^{-5}$ level across the full LISA band, statistically indistinguishable from zero. Even the 95\% band is essentially flat, indicating that no realisation in the ensemble produces a detectable WO signature at this mass. The result at $m_\chi = 10\,\mathrm{keV}$ is qualitatively similar, although a weak signal begins to emerge in the 95\% tail at the $\lesssim 0.6\%$ level. Notably, for the distribution at $10\,\mathrm{keV}$,  the 95\% band extends substantially further from the median than the 68\% band, reflecting the Poissonian nature of the subhalo count when the expected number of WO objects near the image is $\mathcal{O}(1)$ or fewer.

\begin{table}[h]
\centering
\caption{Half-mode masses and minimum-subhalo-mass sweep ranges for the six benchmark WDM particle masses. The tabulated $\msubmin$ range is $[\Mhm/100,\,100\Mhm]$ intersected with $[10^2, 10^9]\,M_\odot$, and is the interval over which $\msubmin$ is varied in Fig.~\ref{fig:msub}. The upper cap of $10^9\,M_\odot$ is reached only at $6\,\mathrm{keV}$, where $\Mhm$ is largest. This $\mchi$-dependent range is distinct from the fixed $\sim 10^4$-$10^7\,M_\odot$ interval that dominates the signal in the LISA band.}
\label{tab:mhm}
\begin{tabular}{ccc}
\toprule
$\mchi\,[\mathrm{keV}]$ &
$\log_{10}(\Mhm/M_\odot)$ &
$\log_{10}\msubmin$ Range $[M_\odot]$ \\
\midrule
$6$   & $7.62$ & $[5.62,\;9.00]$ \\
$10$  & $6.88$ & $[4.88,\;8.88]$ \\
$20$  & $5.88$ & $[3.88,\;7.88]$ \\
$40$  & $4.87$ &  $[2.87,\;6.87]$ \\
$80$  & $3.87$ &  $[2.00,\;5.87]$ \\
$100$ & $3.55$ &  $[2.00,\;5.55]$ \\
\bottomrule
\end{tabular}
\end{table}

Both results are physically plausible. For $m_\chi = 6$ and $10\,\mathrm{keV}$, the half-mode mass lies at $\Mhm \sim 10^{7.6}$ and $10^{6.9} M_\odot$ respectively (Table~\ref{tab:mhm}), placing the WDM suppression scale at or above the subhalo masses that drive appreciable WO distortions across the LISA band. The mass function below $\Mhm$ is strongly suppressed, leaving few WO subhalos near the image per realisation. The few that survive still carry prompt cusps, since the cusp amplitude is set at their infall mass (Sec.~\ref{sec:cusps}), which for these subhalos lies above the $\approx 0.05\,\Mhm$ formation threshold; the null result at these masses is therefore a consequence of the mass-function suppression alone, which renders the WO signal undetectable for the lightest WDM candidates considered here.

\begin{figure*}
    \centering
    \begin{subfigure}{0.45\textwidth}
        \includegraphics[width=\linewidth]{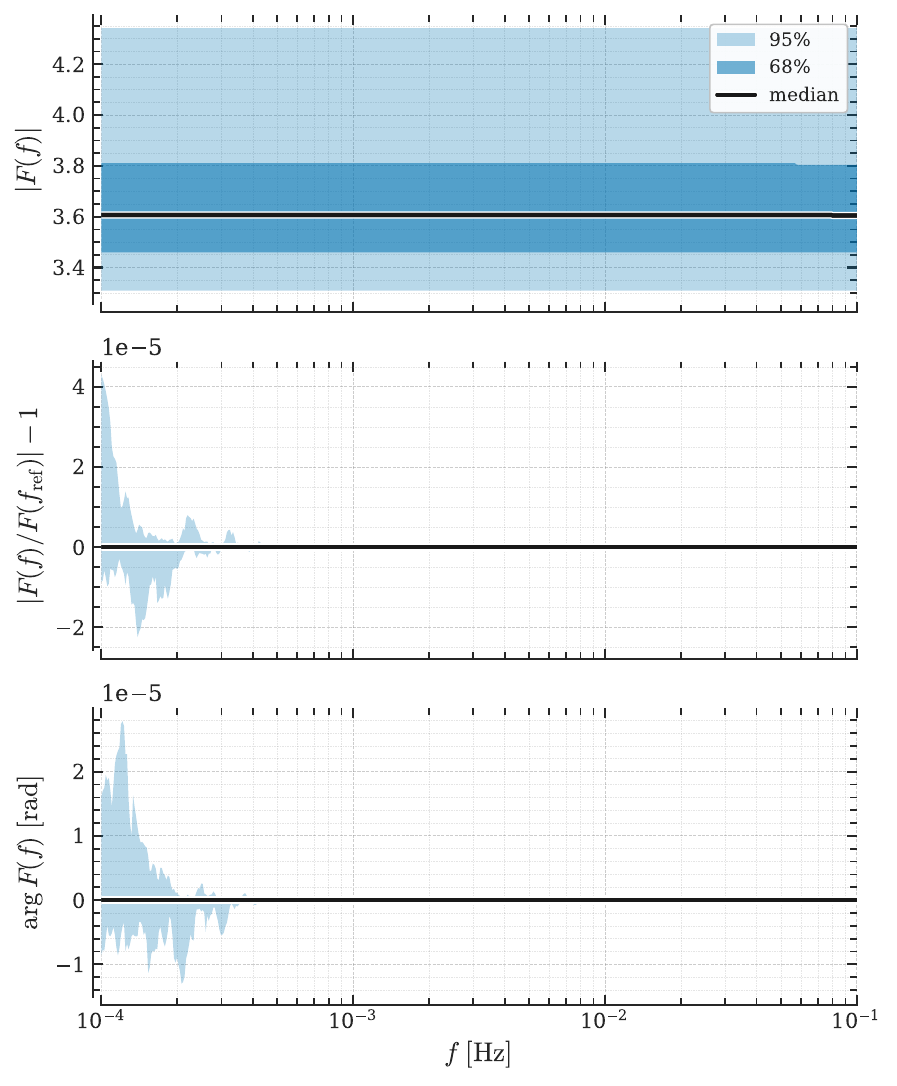}
        \caption{$\mchi = 6\,\mathrm{keV}$}
    \end{subfigure}
    \hfill
    \begin{subfigure}{0.45\textwidth}
        \includegraphics[width=\linewidth]{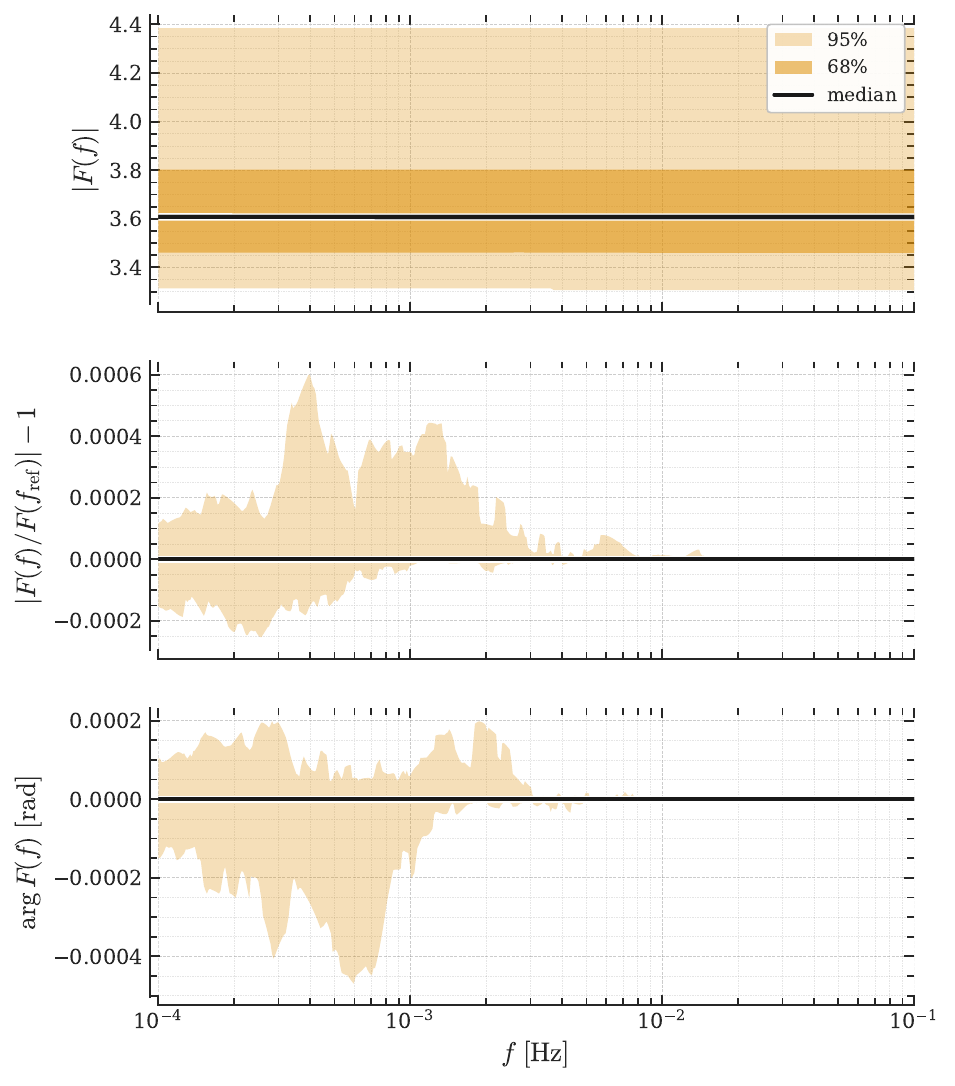}
        \caption{$\mchi = 10\,\mathrm{keV}$}
    \end{subfigure}\\[0.5em]
    \begin{subfigure}{0.45\textwidth}
        \includegraphics[width=\linewidth]{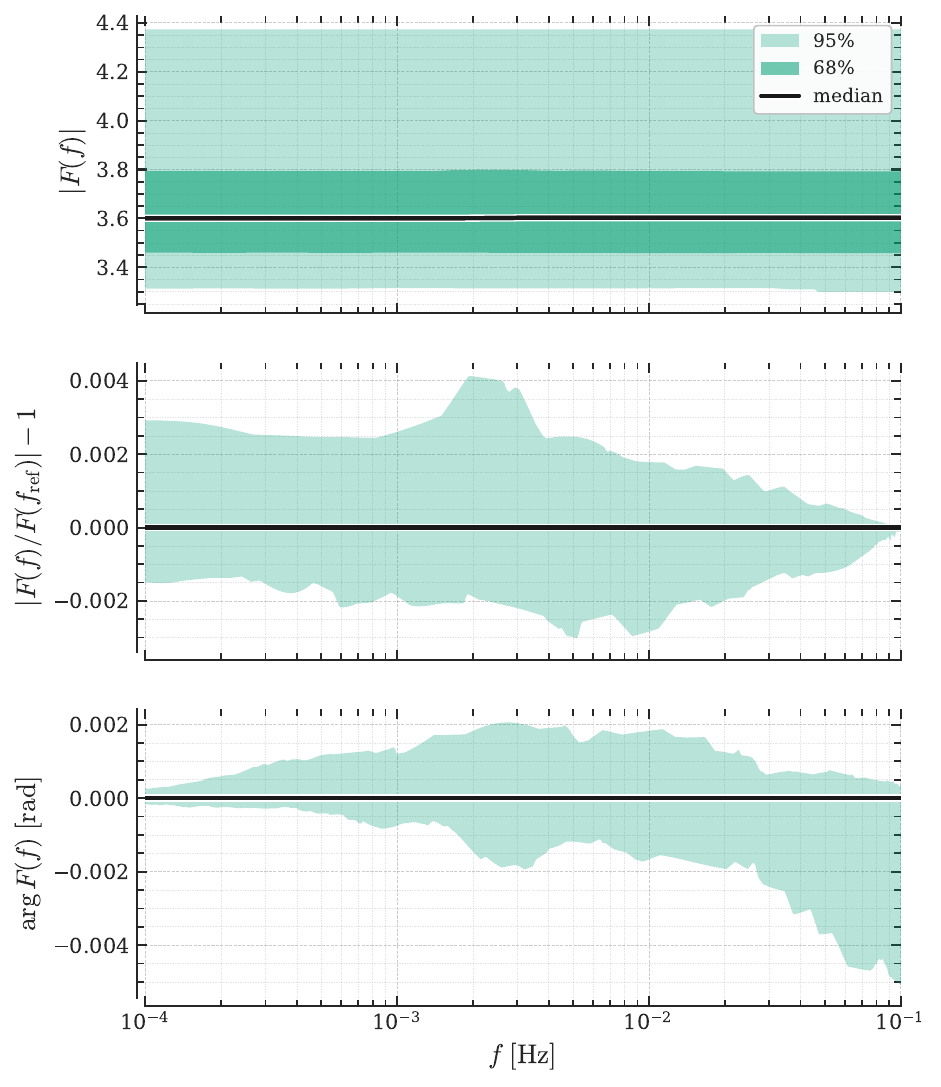}
        \caption{$\mchi = 20\,\mathrm{keV}$}
    \end{subfigure}
    \hfill
    \begin{subfigure}{0.45\textwidth}
        \includegraphics[width=\linewidth]{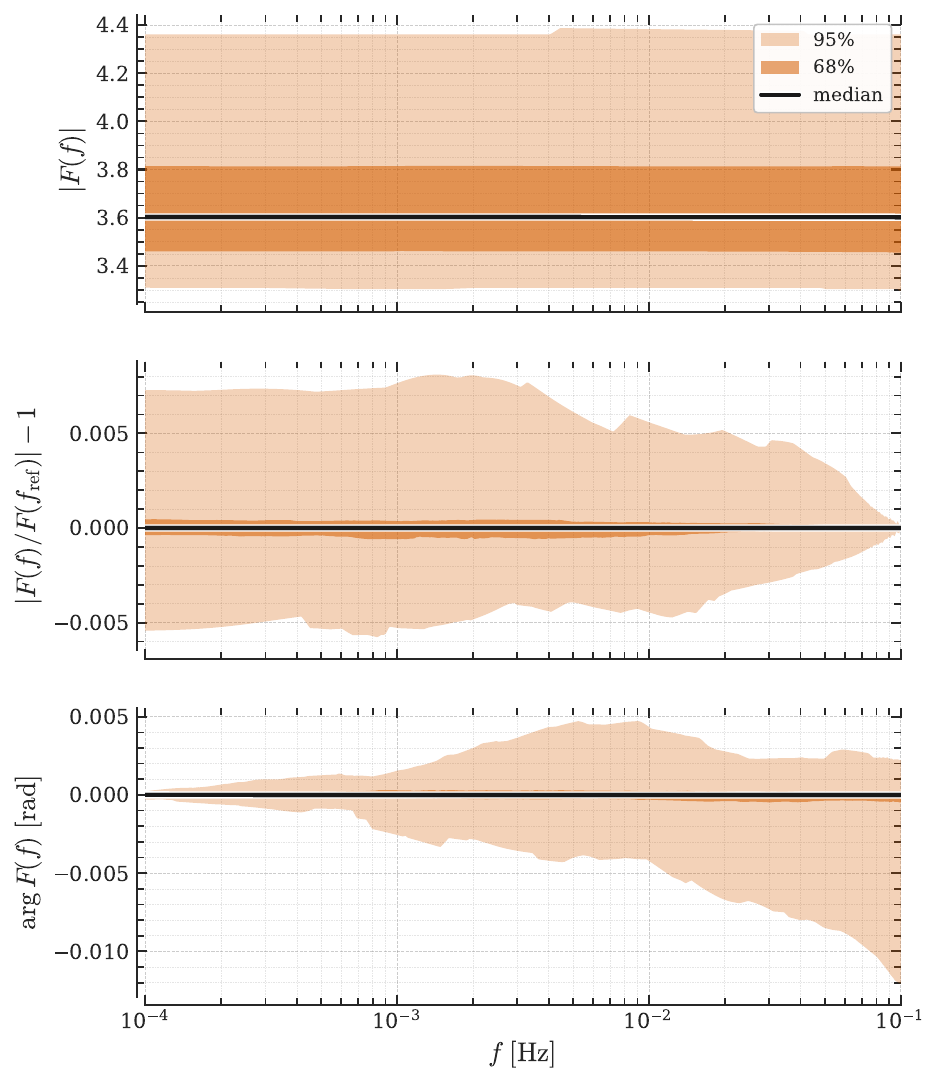}
        \caption{$\mchi = 40\,\mathrm{keV}$}
    \end{subfigure}
    \caption{Frequency-dependent gravitational-wave amplification factor $F(f)$ in the
    WDM model for six values of the WDM particle mass $\mchi$. Each panel shows: absolute amplitude $|F(f)|$ (top); relative amplitude modulation $|F(f)/F(f_{\rm ref})| - 1$ with $f_{\rm ref} = 0.1\,\mathrm{Hz}$ (middle); phase shift $\arg F(f)$ in radians (bottom). Shaded bands show the 68\% (dark) and 95\% (light) percentile ranges over 500
    independent subhalo realisations, where the solid black line is the median. (\textit{Continued below.})}
    \label{fig:ensemble}
\end{figure*}

\begin{figure*}\ContinuedFloat
    \centering
    \begin{subfigure}{0.49\textwidth}
        \includegraphics[width=\linewidth]{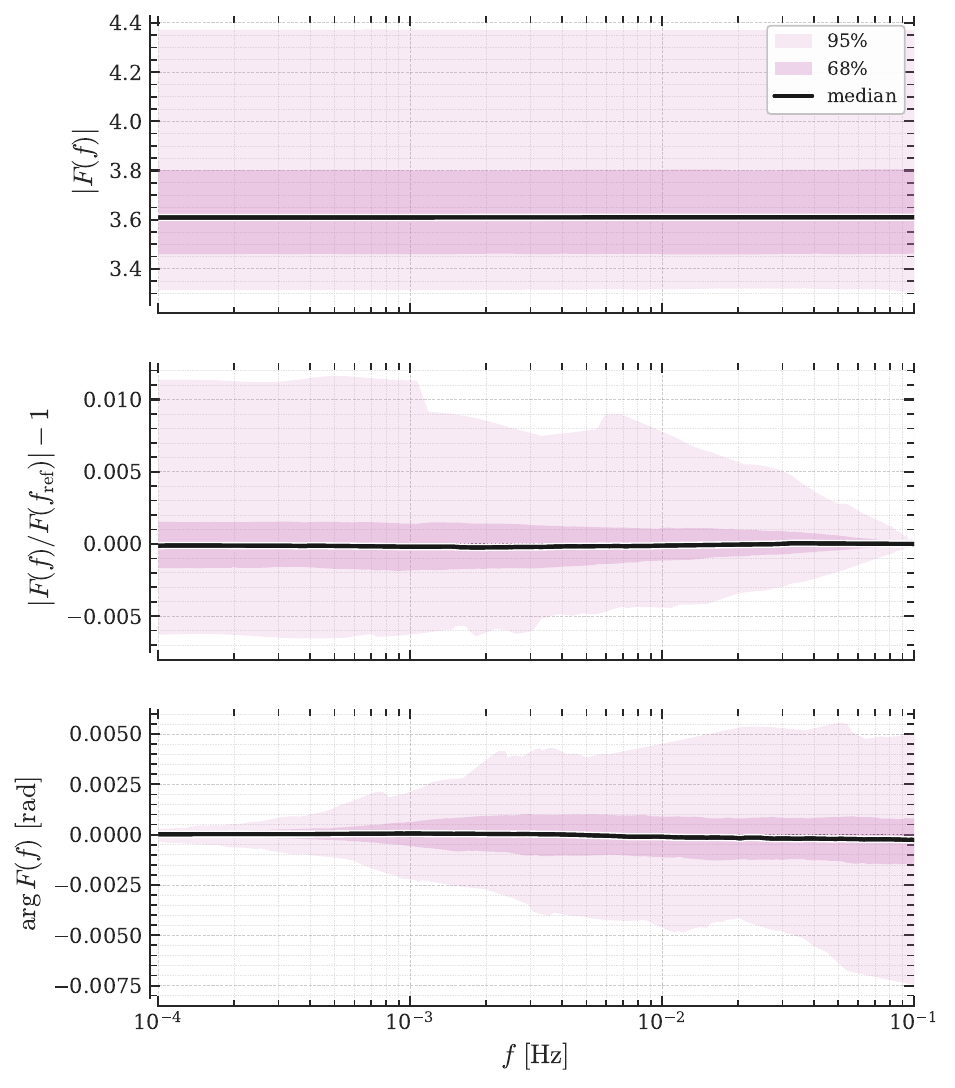}
        \caption{$\mchi = 80\,\mathrm{keV}$}
    \end{subfigure}
    \hfill
    \begin{subfigure}{0.49\textwidth}
        \includegraphics[width=\linewidth]{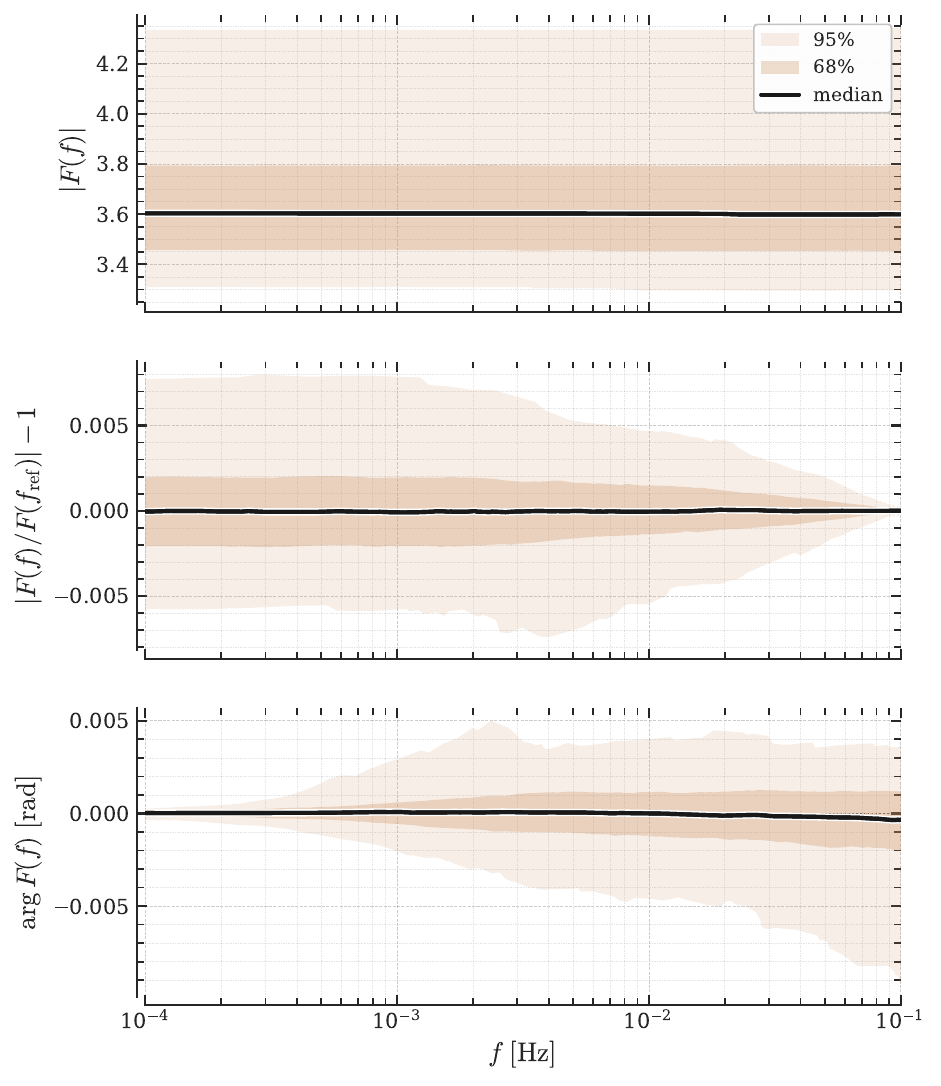}
        \caption{$\mchi = 100\,\mathrm{keV}$}
    \end{subfigure}
    \caption{(\textit{continued}) Ensemble $F(f)$ statistics for
    $\mchi = 80$ and $100\,\mathrm{keV}$.}
\end{figure*}

The modulations become slightly more noticeable at $m_\chi = 20$ and $40 \, \mathrm{keV}$, where $\Mhm$ falls to $\sim 10^{5.9}$ and $10^{4.9}\,M_\odot$ respectively, placing the WDM cutoff within the WO-sensitive mass range for the LISA band. For $m_\chi = 20\,\mathrm{keV}$, the 95\% amplitude band reaches $|F/F_{\rm ref}| - 1 \sim 3\times10^{-3}$ and the phase band $|\arg F| \sim \mathrm{few}\times10^{-3}$ rad, with the signal concentrated between $f \sim 10^{-3}$ and $10^{-2}\,\mathrm{Hz}$, which is the frequency range at which $w \sim 1$ for subhalos near $\Mhm$. A faint oscillatory feature is visible in the median phase at $f \lesssim 10^{-3}\,\mathrm{Hz}$, present consistently across realisations. We identify this as coherent WO modulations among the larger WO subhalos as they transition through $w \sim 1$ at the low-frequency edge of the band. Above $f \sim 10^{-2}\,\mathrm{Hz}$, the phase naturally subsides as those same subhalos enter the geometric-optics regime.

At $m_\chi = 40\,\mathrm{keV}$, the signal is stronger by roughly an order of magnitude in both amplitude and phase, with the 95\% band reaching $|F/F_{\rm ref}|-1 \sim 10^{-2}$ and $|\arg F| \sim 1.5\times 10^{-2}$ rad. The bands are also noticeably wider than at $20\,\mathrm{keV}$, reflecting the larger and more stochastic WO subhalo population. With more objects contributing to the lensing signal, individual realisation-to-realisation variations are amplified. The characteristic peak frequency shifts slightly relative to the $20\, \mathrm{keV}$ case, consistent with the smaller $\Mhm$ driving the dominant signal to higher frequencies.

Beyond $40\,\mathrm{keV}$, the signal appears to saturate. At $m_\chi = 80$ and $100\,\mathrm{keV}$ (as depicted in the corresponding panels of Fig.~\ref{fig:ensemble}) the amplitude and phase distributions are essentially unchanged from the $40\,\mathrm{keV}$ case: the 95\% relative-amplitude band remains roughly at the $\sim 10^{-2}$ level (reaching $\sim 10^{-2}$ at $80\,\mathrm{keV}$ and $\sim 8 \times 10^{-3}$ at $100\,\mathrm{keV}$) and the phase band stays at $\sim\,10^{-3}\,\mathrm{rad}$, with no systematic growth as $\mchi$ increases. By $80\,\mathrm{keV}$ the half-mode mass has fallen to $\Mhm \sim 10^{3.9}\,M_\odot$ (and $\sim 10^{3.6}\,M_\odot$ at $100\,\mathrm{keV}$; Table~\ref{tab:mhm}), slightly below the subhalo mass scales ($\sim 10^4$-$10^7\,M_\odot$). The WDM cutoff therefore no longer suppresses the mass function within the WO-sensitive window, and the relevant subhalo population approaches the CDM abundance shown in Ref.~\cite{Ando2026}. Raising $\mchi$ further therefore adds few WO objects in the mass range that drives the signal. The WO signal thus does not constrain $\mchi$ in the $\gtrsim 40\,\mathrm{keV}$ regime. Such a saturation marks the value of $\mchi$ above which the LISA-band signal becomes indistinguishable from the CDM expectation.

\begin{figure*}
    \centering
    \begin{subfigure}{\textwidth}
        \includegraphics[width=\linewidth]{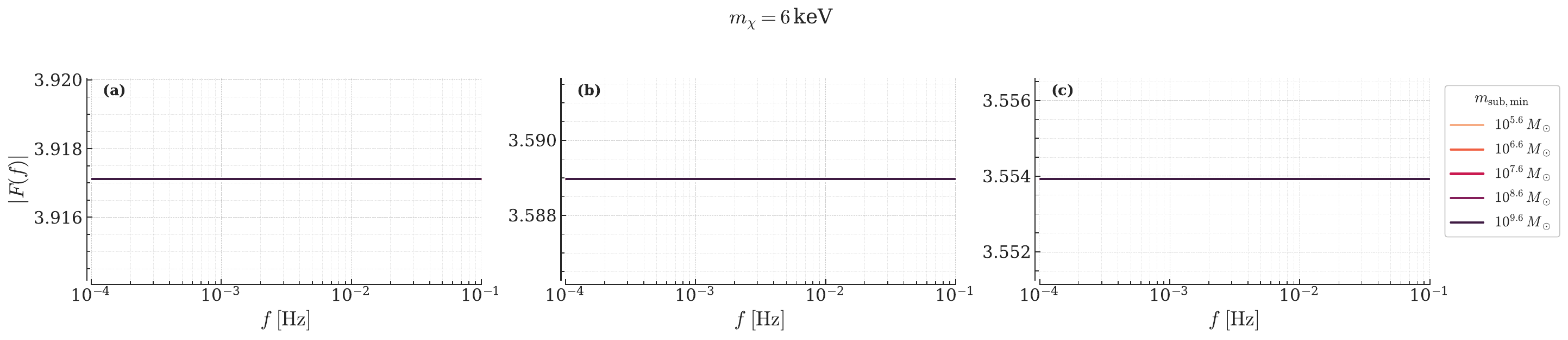}
        \caption*{$\mchi = 6\,\mathrm{keV}$}
    \end{subfigure}\\[0.4em]
    \begin{subfigure}{\textwidth}
        \includegraphics[width=\linewidth]{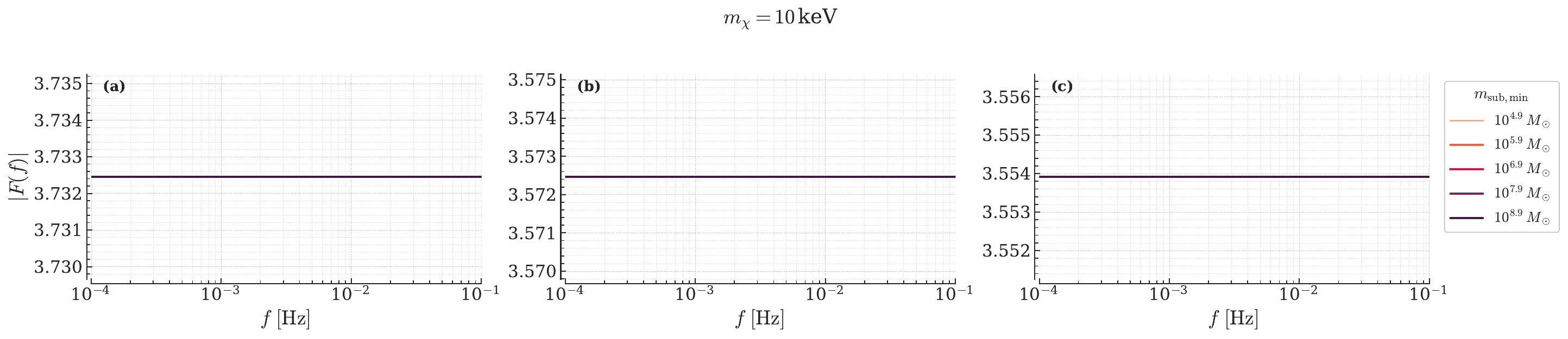}
        \caption*{$\mchi = 10\,\mathrm{keV}$}
    \end{subfigure}\\[0.4em]
    \begin{subfigure}{\textwidth}
        \includegraphics[width=\linewidth]{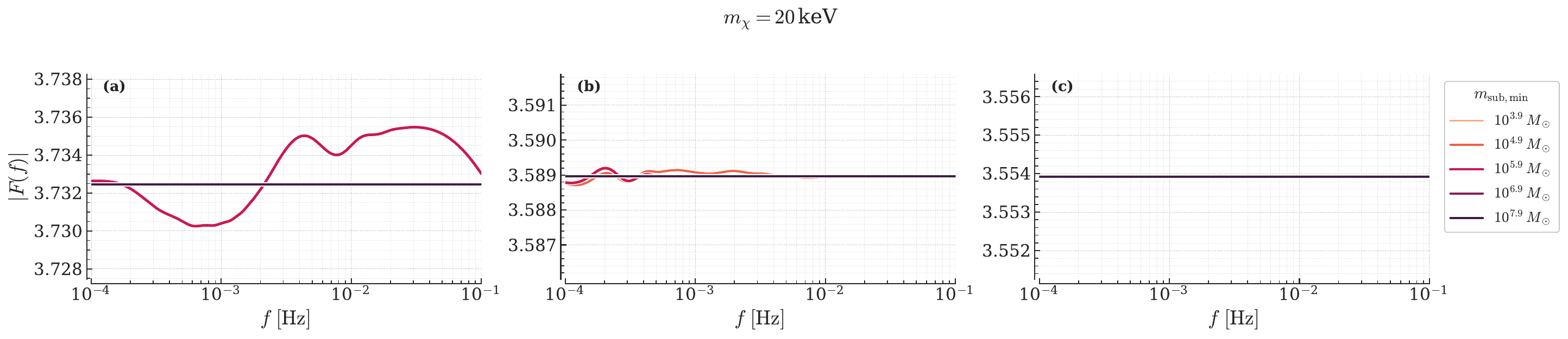}
        \caption*{$\mchi = 20\,\mathrm{keV}$}
    \end{subfigure}
    \caption{Dependence of the GW amplification factor $|F(f)|$ on the minimum subhalo mass $\msubmin$ included in the WO calculation. Each row is labelled by its WDM particle mass $\mchi$, and the three panels (a)-(c) within a row are three independent
realisations at that $\mchi$. (\textit{Continued below.})}
    \label{fig:msub}
\end{figure*}

\begin{figure*}\ContinuedFloat
    \centering
    \begin{subfigure}{\textwidth}
        \includegraphics[width=\linewidth]{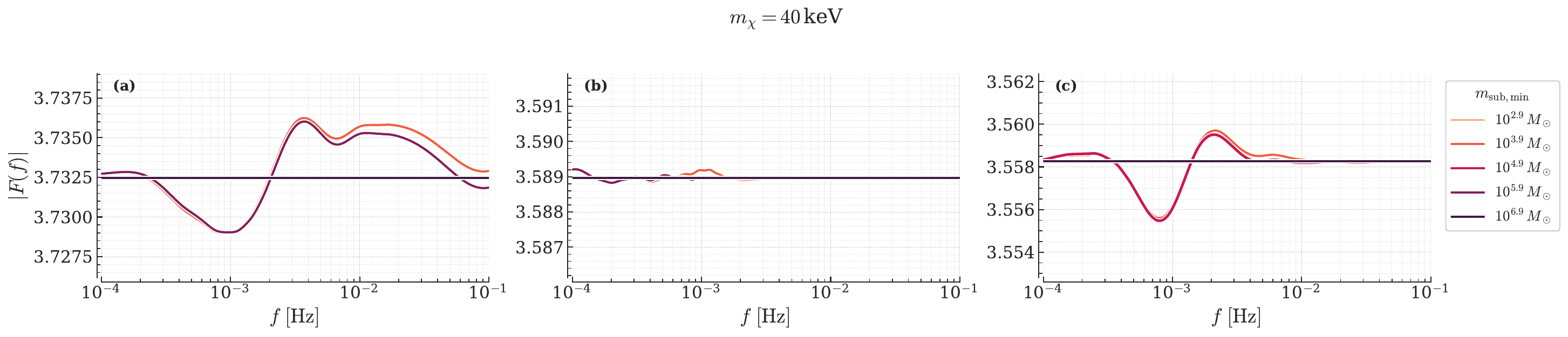}
        \caption*{$\mchi = 40\,\mathrm{keV}$}
    \end{subfigure}\\[0.4em]
    \begin{subfigure}{\textwidth}
        \includegraphics[width=\linewidth]{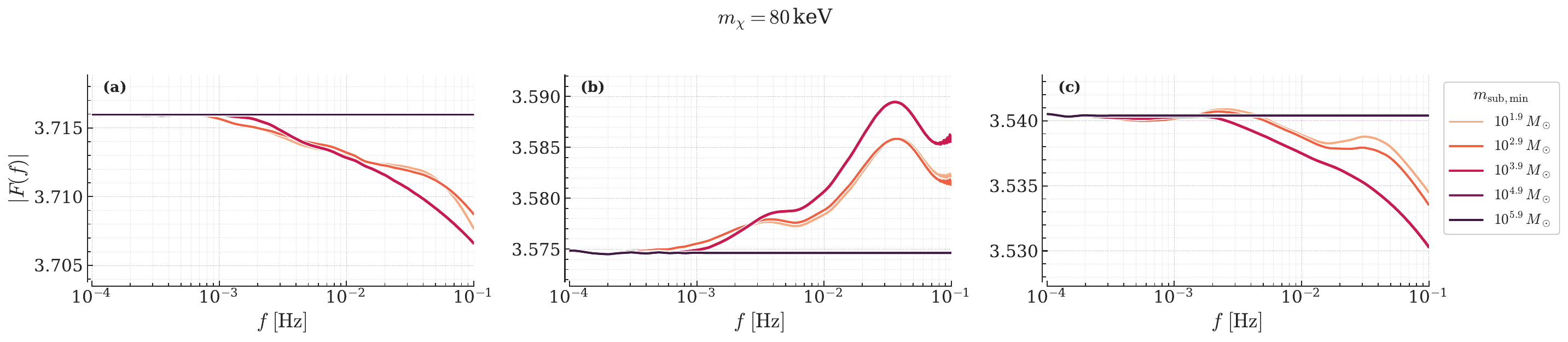}
        \caption*{$\mchi = 80\,\mathrm{keV}$}
    \end{subfigure}\\[0.4em]
    \begin{subfigure}{\textwidth}
        \includegraphics[width=\linewidth]{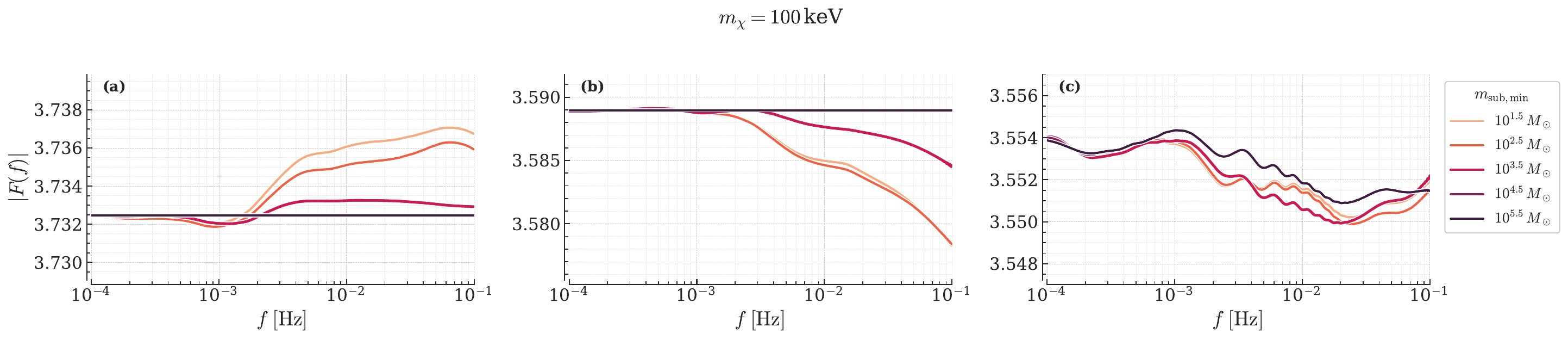}
        \caption*{$\mchi = 100\,\mathrm{keV}$}
    \end{subfigure}
    \caption{(\textit{continued}) WO amplitude sensitivity to $\msubmin$ for
    $\mchi = 40$, $80$, and $100\,\mathrm{keV}$.}
\end{figure*}
To investigate which subhalo mass scales dominate the signal, we vary the minimum subhalo mass $\msubmin$ included in the WO calculation from $10^{-2}\Mhm$ to $10^{2}\Mhm$ in steps of one dex (as shown in Table \ref{tab:mhm}), for three representative realisations at each $m_\chi$. The results are depicted in Fig.~\ref{fig:msub}.

At $m_\chi = 6$ and $10\,\mathrm{keV}$, all five $\msubmin$ curves are near-coincident across every realisation, and lowering the mass threshold by four orders of magnitude produces no measurable change in $|F(f)|$.  We also notice similarly flat trends in Ref.~\cite{Ando2026} for the masses probed. 

At $m_\chi = 20\,\mathrm{keV}$, we see a weak WO signal with low levels of modulation, (panel (a) of the $20,\mathrm{keV}$ row) and only when $\msubmin$ is lowered to $\lesssim 0.1\,\Mhm$. The two lowest $\msubmin$ curves develop a visible oscillatory feature around $f \sim 10^{-3}$--$10^{-2}\,\mathrm{Hz}$, while the three curves with $\msubmin \geq \Mhm$ remain flat. The other two realisations (panels b and c) show no sensitivity at any $\msubmin$, reflecting the stochastic nature of the sub population at this mass. 

For the $m_\chi = 40\,\mathrm{keV}$ case, we find that panels (a) and (c) show clear and consistent $\msubmin$ dependence, and lowering the threshold from $\Mhm$ to $0.1\,\Mhm$ produces a measurable increase in the oscillation amplitude, while reducing further to $0.01\,\Mhm$ adds comparatively little additional signal. This saturation reflects the severe suppression of the WDM mass function below $\sim 0.1\,\Mhm$, which leaves very few subhalos in this range, together with the declining cusp amplitude $A$ as the suppressed linear power spectrum seeds progressively weaker initial overdensities.\footnote{To illustrate, for $m_\chi = 20\,\mathrm{keV}$ at a representative accretion redshift $z_{\rm acc} = 2$, the cusp-halo relation of \cite{delos2025cusp} gives $A \approx 1.1\times10^5\,M_\odot\,\mathrm{kpc}^{-3/2}$ at $0.1\,\Mhm$, rising to $A \approx 3.9\times10^5$ at $\Mhm$ and $A \approx 9.4\times10^5\,M_\odot\,\mathrm{kpc}^{-3/2}$ at $10\,\Mhm$, a factor of $\sim 9$ increase across two decades in mass. At $m_\chi = 40\,\mathrm{keV}$ the same trend holds, with $A$ rising from $3.4\times10^4$ at $0.1\,\Mhm$ to $1.4\times10^5$ at $\Mhm$ and $3.4\times10^5\,M_\odot\,\mathrm{kpc}^{-3/2}$ at $10\,\Mhm$. } The third realisation (panel (b) of the same row) remains flat at all $\msubmin$, indicating seed draws in which no WO subhalo was placed near the image.

At $m_\chi = 80$ and $100\,\mathrm{keV}$, the same saturation seen in the ensemble statistics is apparent in the $\msubmin$ sweep (Fig.~\ref{fig:msub}, $80$ and $100\,\mathrm{keV}$ rows). The absolute variation in $|F(f)|$ across the band is small at the $\sim 0.5\%$ level for individual realisations, and lowering $\msubmin$ below $\sim\Mhm$ produces negligible additional change: the three lowest-$\msubmin$ curves are nearly coincident, while the curves with $\msubmin \gtrsim 10\,\Mhm$ remain flat. Because $\Mhm$ now sits far below the WO-sensitive window, the dominant contribution comes from the fixed $\sim 10^4$--$10^7\,M_\odot$ range set by the LISA frequency band rather than from $\Mhm$ itself, so further lowering the cutoff toward $\Mhm$ no longer adds subhalos in the signal-carrying range. By $m_\chi = 100\,\mathrm{keV}$ the residual $\msubmin$ ordering is no longer evident (panel (c) of the $100,\mathrm{keV}$ row), with the curves overlapping to within their oscillation amplitude. This confirms that the WO signal has reached its CDM-like asymptote. Once the WDM cutoff drops below the LISA-band WO window, neither raising $\mchi$ nor lowering $\msubmin$ alters the lensing signal.

\subsection{Prompt-Cusp Enhancement of the Amplitude Modulation}
\label{sec:cuspimpact}
\begin{figure}
    \centering
    \includegraphics[width=\linewidth]{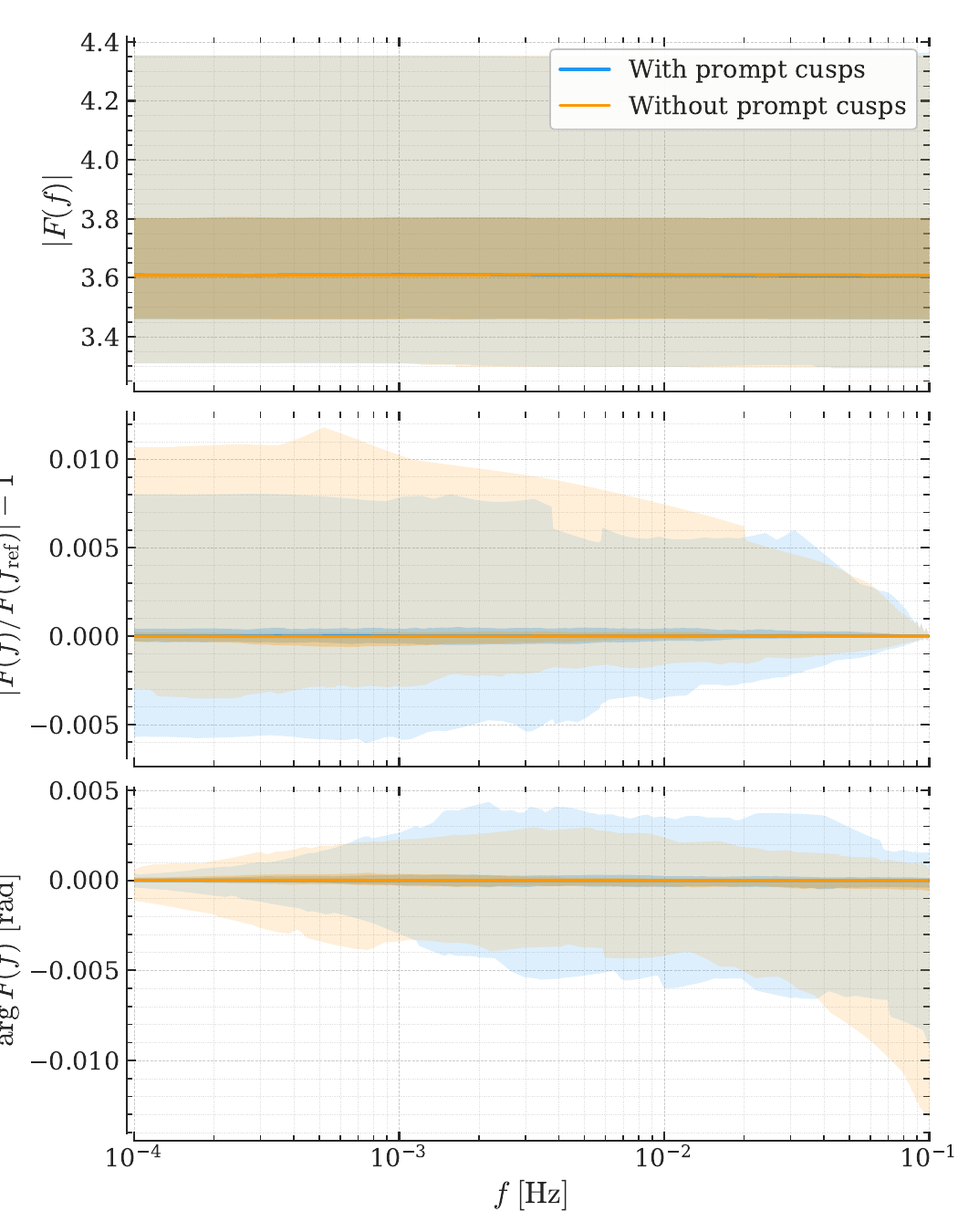}
    \caption{Effect of prompt cusps on the WO signal at $\mchi = 40\,\mathrm{keV}$, over 500
    matched realisations. The two ensembles use identical macrolens configurations, subhalo
    catalogs, and random seeds, differing only in the presence of the $\rho \propto r^{-3/2}$
    central cusp. Blue denotes the fiducial calculation with prompt cusps and orange the
    cusp-free case ($A = 0$). Rows show $|F(f)|$ (top), the relative amplitude modulation
    $|F(f)/F(f_{\rm ref})|-1$ (middle), and the phase $\arg F(f)$ (bottom). Shaded bands are the
    68\% and 95\% percentile ranges and solid lines the medians. The two distributions are
    consistent at the level of the marginal percentile bands; the cusp contribution is recovered
    instead from the matched-pair statistic discussed in the text.}
    \label{fig:cuspcomp}
\end{figure}

We now assess how much of the WO signal is attributable to the prompt cusps. Throughout the preceding analysis we have assumed the presence of a central prompt cusp in every subhalo, fixing its amplitude through the cusp-halo relation of Ref.~\cite{delos2025cusp}. Since this is an intentional modelling choice, it is important to assess how much of the modulation is genuinely sourced by the cusps as opposed to the truncated-NFW subhalos that would be present regardless of the presence of a prompt cusp. We address this by contrasting our fiducial ensemble with an otherwise identical calculation in which the cusp amplitude is set to zero, isolating the cusp contribution and establishing a cusp-free floor for the signal.
\\
To isolate the contribution of the prompt cusps, we repeat the $\mchi = 40\,\mathrm{keV}$ ensemble of Sec.~\ref{sec:ensemble} with the cusp amplitude set to zero ($A = 0$) for every
subhalo, retaining only the NFW component. Both runs use identical macrolens configurations, subhalo catalogs, and random seeds, so that the only difference between them is the presence of the $\rho \propto r^{-3/2}$ central cusp in the WO lens potential. Figure~\ref{fig:cuspcomp} compares the two resulting distributions.

The cusps leave the amplification distribution unchanged, the median remaining at $|F| \simeq 3.6$ with the $68\%$ and $95\%$ bands of the absolute amplitude indistinguishable between the two runs, and the median modulation consistent with zero, so that the useful signal resides entirely in the tails of the distribution, along the lines of sight where a subhalo happens to lie close to the image. Over $500$ realisations the $68\%$ and $95\%$ percentile bands of $|F(f)/F(f_{\rm ref})| - 1$ are consistent between the two runs, agreeing to within their realisation-to-realisation scatter across the band. Because the two runs share the same seeds, the realisations form matched pairs, and differencing them removes the line-of-sight alignment scatter that dominates these marginal distributions. We define $R \equiv s_{\rm cusp}/s_{\rm no\text{-}cusp}$ as the ratio of the per-realisation modulation strength with prompt cusps to that without, evaluated on a common seed so that each matched pair differs only in whether the cusp is present. Because $R$ is broadly distributed, spanning a factor of roughly $40$ between its $5$th and $95$th percentiles, we summarise it on a logarithmic scale. The prompt cusps produce a mild but statistically robust net enhancement, with a geometric-mean ratio of $1.3$ (bootstrap $95\%$ confidence interval $[1.21,\,1.48]$, and a Wilcoxon signed-rank test~\cite{wilcoxon1945individual, hollander2013nonparametric} on $\log R$ giving $p \sim 10^{-5}$). This value is stable under trimming of the tails and is essentially independent of whether the strength is measured by the peak, the $95$th percentile, or the root-mean-square of the modulation. The enhancement is largest where it is physically relevant, reaching $\simeq 1.4$ across the $90$th to $99$th percentiles of the modulation distribution, so that even along the louder realisations that dominate detectability the prompt cusp raises the signal by only a few tens of per cent. It is furthermore accompanied by large scatter, the cusp increasing the modulation on $55\%$ of lines of sight and reducing it on the remainder as the added central mass shifts the interference pattern in either direction. No individual measured modulation can therefore be ascribed to the presence or absence of a prompt cusp, and the cusped and cusp-free signals remain closely similar wherever the modulation is strong enough to be observed. The origin of this similarity, and in particular whether it reflects finite sampling or a true insensitivity of the WO observable to the inner subhalo profile, is examined in Sec.~\ref{sec:whyagree}.

\section{Discussion}
\label{sec:discussion}
\subsection{Assessing the Similarity Between the Cusped and Cusp-Free Signals}
\label{sec:whyagree}
We now probe whether the agreement between the two ensembles found in Sec.~\ref{sec:cuspimpact} is a statistical artefact of finite sampling, or whether it reflects a genuine physical phenomenon that contributes to the insensitivity of the observable. This in  Sec.~\ref{sec:cuspimpact} is at first sight surprising, since prompt cusps are known to dominate the inner structure of low-mass halos \cite{delos2025cusp, Delos2025_WDM}.

It is important to note here that the WO signal is not sourced uniformly across the subhalo population, and that the quantity to which it responds is not the one the prompt cusp most strongly modifies. As established by the $\msubmin$ sweep of Sec.~\ref{sec:ensemble}, the modulation is dominated by subhalos in the range $\msub \sim 10^{4}$--$10^{7}\,M_\odot$, for which the Fresnel radius $R_F = [c\,\deff/(2\pi f)]^{1/2}$ becomes comparable to the subhalo scale radius at different points within the LISA band. This range is set by the detector and does not change with $\mchi$.

Within such a subhalo, the prompt cusp occupies only the innermost region. The dimensionless cusp amplitude of Eq.~(\ref{eq:cusp_nfw}) fixes the radius at which the $\rho \propto r^{-3/2}$ profile gives way to the NFW form, the transition occurring at $r \simeq y^2 r_s$. For our WO population we find a median $y \simeq 0.5$ and $r_s \sim 30\,\mathrm{pc}$, so the cusp governs the profile only within $r \lesssim 8\,\mathrm{pc}$. The Fresnel radius over the LISA band, by contrast, ranges from $\simeq 87\,\mathrm{pc}$ at $f = 10^{-4}\,\mathrm{Hz}$ to $\simeq 2.8\,\mathrm{pc}$ at $f = 10^{-1}\,\mathrm{Hz}$, and the diffraction integral averages the lens potential over a region of this size. The amplification factor therefore responds to the projected mass enclosed within a radius of order $R_F$, rather than to the central density, and the cusp-dominated region lies well inside this radius across most of the band. This is a consequence of the deflection produced at a given impact parameter depending only on the projected mass interior to it, so that a steepening of the profile confined to smaller radii changes the deflection only through the mass the cusp adds, and has no further effect on the deflection.

\begin{figure}[t]
    \centering
    \includegraphics[width=\linewidth]{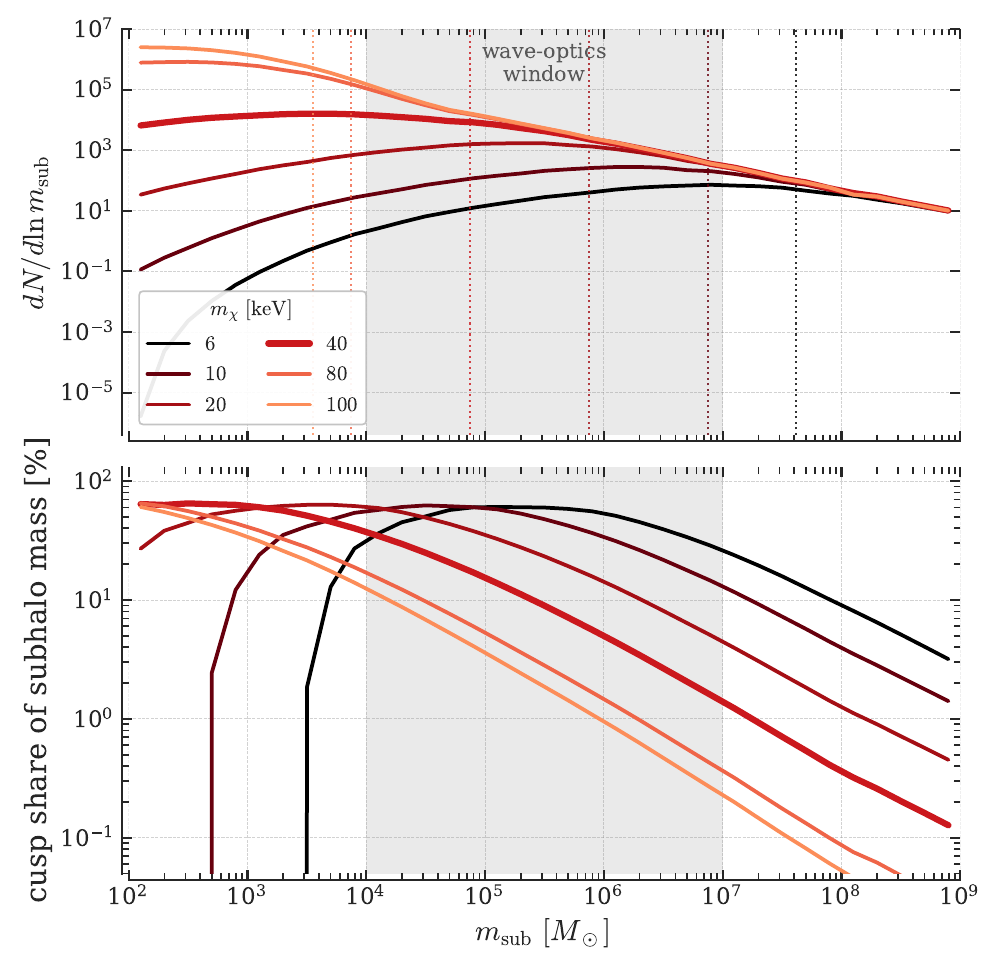}
    \caption{Subhalo abundance and prompt-cusp mass fraction versus subhalo mass. \textit{Top}: abundance of subhalos per logarithmic mass interval for each WDM particle mass. The dotted vertical lines mark the corresponding $\Mhm$. \textit{Bottom}: fraction of the bound subhalo mass carried by the prompt cusp, a structural indicator. The projected mass within $R_F$ that sources the modulation is larger than this share (see text). The shaded band marks the $\msub \sim 10^{4}$-$10^{7}\, M_\odot$ range that dominates the LISA-band signal (Sec.~\ref{sec:ensemble}), which is set by the detector and independent of $\mchi$.  Within the shaded band the cusp supplies up to $\simeq 40\%$ of the subhalo mass only for the smallest $\mchi$, for which $\Mhm$ has already suppressed the abundance there, and falls to a few per cent by $100\,\mathrm{keV}$. Each curve drops to zero at low present-day mass where the infall-mass formation threshold ($\approx 0.05\,\Mhm$; Sec.~\ref{sec:cusps}) is crossed. Since the amplitude is set at infall, this drop occurs well below $0.05\,\Mhm$ in present-day mass.}
    \label{fig:cuspwindow}
\end{figure}

Figure~\ref{fig:cuspwindow} quantifies how much of the subhalo the prompt cusp supplies. The lower panel gives the cusp share of the bound subhalo mass as a function of subhalo mass for each $\mchi$. At fixed subhalo mass the share falls steeply with increasing $\mchi$, since a smaller free-streaming scale yields less massive cusps, in agreement with the behaviour reported in Ref.~\cite{delos2025cusp}. Within the signal-carrying range, at $10^{5}\,M_\odot$ the share drops from $\simeq 38\%$ at $\mchi = 6\,\mathrm{keV}$ to $\simeq 13\%$ at $\mchi = 40\,\mathrm{keV}$ and $\simeq 4\%$ at $100\,\mathrm{keV}$.

While the mass share is an indicator of the structure of the subhalo, the quantity that affects the modulation is the projected mass enclosed within $\sim R_F$, to which the centrally concentrated cusp contributes a larger fraction. At $10^{5}\,M_\odot$ and $\mchi = 40\,\mathrm{keV}$ this projected share is $\simeq 20\%$, roughly $1.5$ times the overall bound-mass value. A boost of this size to a percent-level modulation accounts for the geometric-mean enhancement $R \simeq 1.3$ found in Sec.~\ref{sec:cuspimpact}. The prompt cusp therefore raises a percent-level modulation by only a few tens of per cent, leaving it at the same order and producing no distinctive signature.

The upper panel of Fig.~\ref{fig:cuspwindow} accounts for the remaining $\mchi$ dependence. The abundance of subhalos within the signal-carrying range is governed by $\Mhm$, which must fall below that range for the population to be present. Counting the subhalos available per realisation there, we find that a substantial number of subhalos become populated only above $20\,\mathrm{keV}$, which is the origin of the null results at $6$ and $10\,\mathrm{keV}$ reported in Sec.~\ref{sec:ensemble}. The particle masses at which prompt cusps make up the largest fraction of the subhalo mass are therefore also those at which the WO window is most severely depopulated. Across the WDM parameter space the conclusion holds at both ends of the range considered. Where prompt cusps make up as much as $\simeq 40\%$ of the subhalo mass the signal is absent, and where the signal is present their imprint on the modulation is modest for the reason above. In neither regime do they produce a distinctive amplitude-modulation signature that separates cleanly from that of truncated-NFW subhalos alone.

The realisations that LISA could plausibly observe are those in the well-aligned tail, where both cases predict percent-level modulations. A measured WO distortion would therefore establish the presence of substructure on Fresnel scales, without constraining whether those subhalos carry prompt cusps.

\subsection{Detectability}
Let $(\mathrm{S/N})_0$ denote the intrinsic signal-to-noise ratio of the unlensed GW source. As noted by Ref.~\cite{Ando2026}, for a strongly magnified event with amplification $|F| \sim 3$-$5$, the observed $\mathrm{S/N}$ scales as $|F|\,(\mathrm{S/N})_0$.  A fractional amplitude perturbation $\delta h/h$ is detectable when $(\delta h/h)|F|(\mathrm{S/N})_0 \gtrsim 1$, while phase perturbations become measurable once $\delta\phi \gtrsim [|F|(\mathrm{S/N})_0
]^{-1}$.  For our $m_\chi = 40\,\mathrm{keV}$ fiducial, the WO amplitude and phase modulations reach $\sim 10^{-2}$ in the outer 95\% tail of the realisation distribution, with typical (median) realisations lying well below this value. For a strongly lensed event with $|F| \sim 3.6$ and intrinsic $(\mathrm{S/N})_0 \gtrsim 100$, a tail modulation of $\sim 10^{-2}$ yields $(\delta h/h)\,|F|\,(\mathrm{S/N})_0 \sim 4$, a several-$\sigma$ effect; for the loudest massive black-hole binary events in the LISA band with $(\mathrm{S/N})_0 \sim 10^3$~\citep{amaro2017laser}, the same modulation reaches $\sim 40$. The detection significance is therefore strongly realisation-dependent. Percent-level WO distortions from $40\,\mathrm{keV}$ WDM subhalos are detectable at high statistical significance in the favorable tail along the line of sight, and especially for the loudest LISA sources, whereas typical realisations fall below the detection threshold at $(\mathrm{S/N})_0 \sim 100$. We stress that this estimate refers to the detectability of the total WO imprint of the subhalo population, and not to a measurement of the prompt-cusp contribution specifically. As shown in Sec.~\ref{sec:cuspimpact}, the predicted modulation distributions with and without cusps overlap almost entirely, so that a single detection of a percent-level distortion would establish the presence of WO-scale substructure without by itself discriminating between cusped and cusp-free inner profiles.

\section{Conclusion}
\label{sec:conclusion}
We have computed the WO amplification factor $F(f)$ for strongly lensed gravitational waves propagating through a dark-matter host halo populated with WDM subhalos and prompt-cusp inner profiles. In the LISA band we find percent-level modulations in $F(f)$ once the WDM particle mass reaches $\mchi \gtrsim 40\,\mathrm{keV}$. For $\mchi \lesssim 10\,\mathrm{keV}$ the half-mode mass $\Mhm$ lies at or above the upper edge of this window, suppressing virtually all WO subhalos and leaving the signal undetectable. As $\mchi$ increases from $\sim 20\,\mathrm{keV}$, $\Mhm$ enters the window and the signal grows, reaching amplitude and phase modulations of order $10^{-2}$ by $40\,\mathrm{keV}$; above this it saturates, with $\mchi = 80$ and $100\,\mathrm{keV}$ holding at the same $\sim 10^{-2}$ level as $\Mhm$ drops below the window and the LISA-band population approaches its CDM abundance. The WO signal can therefore probe $\mchi$ only in the intermediate $\sim 20$--$40\,\mathrm{keV}$ range, saturating to a CDM-like plateau above it. Repeating the $40\,\mathrm{keV}$ ensemble with the cusp amplitude set to zero on identical subhalo catalogs shows that prompt cusps enhance the modulation only modestly: the marginal percentile bands of the two cases are consistent, and a matched-pair analysis over $500$ realisations gives a net enhancement of $R \simeq 1.3$, rising to $\simeq 1.4$ along the loudest modulations. The bulk of the signal is thus sourced by the truncated-NFW subhalo population itself, and this near-degeneracy is robust across the WDM models we consider rather than specific to the fiducial one. The reason is a mismatch of scales: at the particle masses where prompt cusps make up the largest fraction of the subhalo mass, free-streaming has already depopulated the WO window, while at the masses where the window is populated the cusp is centrally concentrated on scales to which the modulation is largely insensitive. Prompt cusps therefore leave no distinctive imprint separating them from cusp-free subhalos in the LISA-band signal. Strongly lensed LISA gravitational waves may probe the abundance of substructure on Fresnel scales, but, in the WDM models and lens configuration examined here, do not provide a clean diagnostic of prompt-cusp inner profiles.

\acknowledgments
The work of SA was supported by Grant-in-Aid for Scientific Research from the JSPS KAKENHI grant number JP24K07039. BT acknowledges support from the \textsc{infn} Research Grant TAsP (Theoretical Astroparticle Physics), as well as support from the PNRR grant "ex DM 118" scholarship and the European Union – Next Generation EU.

\bibliographystyle{apsrev4-2}
\bibliography{refs}   

@article{villarrubia2025gravitational,
  author        = {Villarrubia-Rojo, H. and Savastano, S. and Zumalac{\'a}rregui, M. and Choi, L. and Goyal, S. and Dai, L. and Tambalo, G.},
  title         = {{Gravitational lensing of waves: novel methods for wave-optics phenomena}},
  journal       = {Phys. Rev. D},
  volume        = {111},
  pages         = {103539},
  year          = {2025},
  doi           = {10.1103/PhysRevD.111.103539},
  eprint        = {2409.04606},
  archivePrefix = {arXiv},
  primaryClass  = {gr-qc}
}

@article{Ando2026,
  author        = {Ando, S.},
  title         = {{Wave-optics imprints of dark matter subhalos on strongly lensed gravitational waves}},
  journal       = {},
  year          = {2026},
  eprint        = {2603.04267},
  archivePrefix = {arXiv},
  primaryClass  = {astro-ph.CO}
}

@article{Guo2022,
  author        = {Guo, X. and Lu, Y.},
  title         = {{Probing the nature of dark matter via gravitational waves lensed by small dark matter halos}},
  journal       = {Phys. Rev. D},
  volume        = {106},
  pages         = {023018},
  year          = {2022},
  doi           = {10.1103/PhysRevD.106.023018},
  eprint        = {2207.00325},
  archivePrefix = {arXiv},
  primaryClass  = {astro-ph.CO}
}

@article{Oguri2018,
  author        = {Oguri, M.},
  title         = {{Effect of gravitational lensing on the distribution of gravitational waves from distant binary black hole mergers}},
  journal       = {Mon. Not. R. Astron. Soc.},
  volume        = {480},
  pages         = {3842},
  year          = {2018},
  doi           = {10.1093/mnras/sty2145},
  eprint        = {1807.02584},
  archivePrefix = {arXiv},
  primaryClass  = {astro-ph.CO}
}

@article{Fairbairn2023,
  author        = {Fairbairn, M. and Urrutia, J. and Vaskonen, V.},
  title         = {{Microlensing of gravitational waves by dark matter structures}},
  journal       = {J. Cosmol. Astropart. Phys.},
  volume        = {2023},
  number        = {07},
  pages         = {007},
  year          = {2023},
  doi           = {10.1088/1475-7516/2023/07/007},
  eprint        = {2210.13436},
  archivePrefix = {arXiv},
  primaryClass  = {astro-ph.CO}
}

@article{NFW1996,
  author        = {Navarro, J. F. and Frenk, C. S. and White, S. D. M.},
  title         = {{The structure of cold dark matter halos}},
  journal       = {Astrophys. J.},
  volume        = {462},
  pages         = {563},
  year          = {1996},
  doi           = {10.1086/177173},
  eprint        = {astro-ph/9508025},
  archivePrefix = {arXiv}
}

@article{correa2015accretion,
  author        = {Correa, C. A. and Wyithe, J. S. B. and Schaye, J. and Duffy, A. R.},
  title         = {{The accretion history of dark matter haloes -- III. A physical model for the concentration--mass relation}},
  journal       = {Mon. Not. R. Astron. Soc.},
  volume        = {452},
  pages         = {1217},
  year          = {2015},
  doi           = {10.1093/mnras/stv1363},
  eprint        = {1502.00391},
  archivePrefix = {arXiv},
  primaryClass  = {astro-ph.CO}
}

@article{dekker2022warm,
  author        = {Dekker, A. and Ando, S. and Correa, C. A. and Ng, K. C. Y.},
  title         = {{Warm dark matter constraints using Milky Way satellite observations and subhalo evolution modeling}},
  journal       = {Phys. Rev. D},
  volume        = {106},
  pages         = {123026},
  year          = {2022},
  doi           = {10.1103/PhysRevD.106.123026},
  eprint        = {2111.13137},
  archivePrefix = {arXiv},
  primaryClass  = {astro-ph.CO}
}

@article{hiroshima2018modeling,
  author        = {Hiroshima, N. and Ando, S. and Ishiyama, T.},
  title         = {{Modeling evolution of dark matter substructure and annihilation boost}},
  journal       = {Phys. Rev. D},
  volume        = {97},
  pages         = {123002},
  year          = {2018},
  doi           = {10.1103/PhysRevD.97.123002},
  eprint        = {1803.07691},
  archivePrefix = {arXiv},
  primaryClass  = {astro-ph.CO}
}

@article{ludlow2016mass,
  author        = {Ludlow, A. D. and Bose, S. and Angulo, R. E. and Wang, L. and Hellwing, W. A. and Navarro, J. F. and Cole, S. and Frenk, C. S.},
  title         = {{The mass--concentration--redshift relation of cold and warm dark matter haloes}},
  journal       = {Mon. Not. R. Astron. Soc.},
  volume        = {460},
  pages         = {1214},
  year          = {2016},
  doi           = {10.1093/mnras/stw1046},
  eprint        = {1601.02624},
  archivePrefix = {arXiv},
  primaryClass  = {astro-ph.CO}
}

@article{Vegetti2024,
  author        = {Vegetti, S. and Birrer, S. and Despali, G. and Fassnacht, C. D. and Gilman, D. and Hezaveh, Y. and Perreault Levasseur, L. and McKean, J. P. and Powell, D. M. and O'Riordan, C. M. and Vernardos, G.},
  title         = {{Strong gravitational lensing as a probe of dark matter}},
  journal       = {Space Sci. Rev.},
  volume        = {220},
  pages         = {58},
  year          = {2024},
  doi           = {10.1007/s11214-024-01087-w},
  eprint        = {2306.11781},
  archivePrefix = {arXiv},
  primaryClass  = {astro-ph.CO}
}

@article{Takahashi2003,
  author        = {Takahashi, R. and Nakamura, T.},
  title         = {{Wave effects in the gravitational lensing of gravitational waves from chirping binaries}},
  journal       = {Astrophys. J.},
  volume        = {595},
  pages         = {1039},
  year          = {2003},
  doi           = {10.1086/377430},
  eprint        = {astro-ph/0305055},
  archivePrefix = {arXiv}
}

@article{nadler2021constraints,
  author        = {Nadler, E. O. and Drlica-Wagner, A. and Bechtol, K. and Mau, S. and Wechsler, R. H. and Gluscevic, V. and Boddy, K. and Pace, A. B. and Li, T. S. and McNanna, M. and others},
  collaboration = {DES},
  title         = {{Constraints on dark matter properties from observations of Milky Way satellite galaxies}},
  journal       = {Phys. Rev. Lett.},
  volume        = {126},
  pages         = {091101},
  year          = {2021},
  doi           = {10.1103/PhysRevLett.126.091101},
  eprint        = {2008.00022},
  archivePrefix = {arXiv},
  primaryClass  = {astro-ph.CO}
}

@article{Gilman2020,
  author        = {Gilman, D. and Birrer, S. and Nierenberg, A. and Treu, T. and Du, X. and Benson, A.},
  title         = {{Warm dark matter chills out: constraints on the halo mass function and the free-streaming length of dark matter with eight quadruple-image strong gravitational lenses}},
  journal       = {Mon. Not. R. Astron. Soc.},
  volume        = {491},
  pages         = {6077},
  year          = {2020},
  doi           = {10.1093/mnras/stz3480},
  eprint        = {1908.06983},
  archivePrefix = {arXiv},
  primaryClass  = {astro-ph.CO}
}

@article{sereno2010strong,
  author        = {Sereno, M. and Sesana, A. and Bleuler, A. and Jetzer, Ph. and Volonteri, M. and Begelman, M. C.},
  title         = {{Strong lensing of gravitational waves as seen by LISA}},
  journal       = {Phys. Rev. Lett.},
  volume        = {105},
  pages         = {251101},
  year          = {2010},
  doi           = {10.1103/PhysRevLett.105.251101},
  eprint        = {1011.5238},
  archivePrefix = {arXiv},
  primaryClass  = {astro-ph.CO}
}

@article{gutierrez2025strong,
  author        = {Guti{\'e}rrez, J. and Lagos, M.},
  title         = {{Strong-lensing rates of massive black hole binaries in LISA}},
  journal       = {Phys. Rev. D},
  volume        = {112},
  pages         = {123512},
  year          = {2025},
  doi           = {10.1103/yd5h-ql5f},
  eprint        = {2510.02061},
  archivePrefix = {arXiv},
  primaryClass  = {astro-ph.CO}
}

@article{Viel2005,
  author        = {Viel, M. and Lesgourgues, J. and Haehnelt, M. G. and Matarrese, S. and Riotto, A.},
  title         = {{Constraining warm dark matter candidates including sterile neutrinos and light gravitinos with WMAP and the Lyman-$\alpha$ forest}},
  journal       = {Phys. Rev. D},
  volume        = {71},
  pages         = {063534},
  year          = {2005},
  doi           = {10.1103/PhysRevD.71.063534},
  eprint        = {astro-ph/0501562},
  archivePrefix = {arXiv}
}

@article{Delos2018,
  author        = {Delos, M. S. and Erickcek, A. L. and Bailey, A. P. and Alvarez, M. A.},
  title         = {{Are ultracompact minihalos really ultracompact?}},
  journal       = {Phys. Rev. D},
  volume        = {97},
  pages         = {041303},
  year          = {2018},
  doi           = {10.1103/PhysRevD.97.041303},
  eprint        = {1712.05421},
  archivePrefix = {arXiv},
  primaryClass  = {astro-ph.CO}
}

@article{delos2019predicting,
  author        = {Delos, M. S. and Bruff, M. and Erickcek, A. L.},
  title         = {{Predicting the density profiles of the first halos}},
  journal       = {Phys. Rev. D},
  volume        = {100},
  pages         = {023523},
  year          = {2019},
  doi           = {10.1103/PhysRevD.100.023523},
  eprint        = {1905.05766},
  archivePrefix = {arXiv},
  primaryClass  = {astro-ph.CO}
}

@article{delos2025cusp,
  author        = {Delos, M. S.},
  title         = {{The cusp--halo relation}},
  journal       = {Astrophys. J.},
  volume        = {993},
  pages         = {93},
  year          = {2025},
  doi           = {10.3847/1538-4357/ae0625},
  eprint        = {2506.03240},
  archivePrefix = {arXiv},
  primaryClass  = {astro-ph.CO}
}

@article{brando2025signatures,
  author        = {Brando, G. and Goyal, S. and Savastano, S. and Villarrubia-Rojo, H. and Zumalac{\'a}rregui, M.},
  title         = {{Signatures of dark and baryonic structures on weakly lensed gravitational waves}},
  journal       = {Phys. Rev. D},
  volume        = {111},
  pages         = {024068},
  year          = {2025},
  doi           = {10.1103/PhysRevD.111.024068},
  eprint        = {2407.04052},
  archivePrefix = {arXiv},
  primaryClass  = {gr-qc}
}

@article{Delos2025_WDM,
  author        = {Delos, M. S. and Ahvazi, N. and Benson, A.},
  title         = {{Testing warm dark matter with kinematics of the smallest galaxies}},
  journal       = {},
  year          = {2025},
  eprint        = {2512.04156},
  archivePrefix = {arXiv},
  primaryClass  = {astro-ph.CO}
}

@article{Lovell2014,
  author        = {Lovell, M. R. and Frenk, C. S. and Eke, V. R. and Jenkins, A. and Gao, L. and Theuns, T.},
  title         = {{The properties of warm dark matter haloes}},
  journal       = {Mon. Not. R. Astron. Soc.},
  volume        = {439},
  pages         = {300},
  year          = {2014},
  doi           = {10.1093/mnras/stt2431},
  eprint        = {1308.1399},
  archivePrefix = {arXiv},
  primaryClass  = {astro-ph.CO}
}

@article{Bose2016,
  author        = {Bose, S. and Hellwing, W. A. and Frenk, C. S. and Jenkins, A. and Lovell, M. R. and Helly, J. C. and Li, B.},
  title         = {{The Copernicus Complexio: statistical properties of warm dark matter haloes}},
  journal       = {Mon. Not. R. Astron. Soc.},
  volume        = {455},
  pages         = {318},
  year          = {2016},
  doi           = {10.1093/mnras/stv2294},
  eprint        = {1507.01998},
  archivePrefix = {arXiv},
  primaryClass  = {astro-ph.CO}
}

@article{amaro2017laser,
  author        = {Amaro-Seoane, P. and Audley, H. and Babak, S. and Baker, J. and Barausse, E. and Bender, P. and Berti, E. and Binetruy, P. and Born, M. and Bortoluzzi, D. and others},
  title         = {{Laser Interferometer Space Antenna}},
  journal       = {},
  year          = {2017},
  eprint        = {1702.00786},
  archivePrefix = {arXiv},
  primaryClass  = {astro-ph.IM}
}

@article{Springel2008,
  author        = {Springel, V. and Wang, J. and Vogelsberger, M. and Ludlow, A. and Jenkins, A. and Helmi, A. and Navarro, J. F. and Frenk, C. S. and White, S. D. M.},
  title         = {{The Aquarius Project: the subhaloes of galactic haloes}},
  journal       = {Mon. Not. R. Astron. Soc.},
  volume        = {391},
  pages         = {1685},
  year          = {2008},
  doi           = {10.1111/j.1365-2966.2008.14066.x},
  eprint        = {0809.0898},
  archivePrefix = {arXiv},
  primaryClass  = {astro-ph}
}

@article{Bullock2017,
  author        = {Bullock, J. S. and Boylan-Kolchin, M.},
  title         = {{Small-scale challenges to the $\Lambda$CDM paradigm}},
  journal       = {Annu. Rev. Astron. Astrophys.},
  volume        = {55},
  pages         = {343},
  year          = {2017},
  doi           = {10.1146/annurev-astro-091916-055313},
  eprint        = {1707.04256},
  archivePrefix = {arXiv},
  primaryClass  = {astro-ph.CO}
}

@article{aghanim2020planck,
  author        = {Aghanim, N. and others},
  collaboration = {Planck},
  title         = {{Planck 2018 results. VI. Cosmological parameters}},
  journal       = {Astron. Astrophys.},
  volume        = {641},
  pages         = {A6},
  year          = {2020},
  doi           = {10.1051/0004-6361/201833910},
  eprint        = {1807.06209},
  archivePrefix = {arXiv},
  primaryClass  = {astro-ph.CO}
}

@article{Simon2019,
  author        = {Simon, J. D.},
  title         = {{The faintest dwarf galaxies}},
  journal       = {Annu. Rev. Astron. Astrophys.},
  volume        = {57},
  pages         = {375},
  year          = {2019},
  doi           = {10.1146/annurev-astro-091918-104453},
  eprint        = {1901.05465},
  archivePrefix = {arXiv},
  primaryClass  = {astro-ph.GA}
}

@article{Ando:2026eam,
  author        = {Ando, S.},
  title         = {{Wave-optics imprints of dark matter subhalos on strongly lensed gravitational waves. II. Saddle images and detectability}},
  journal       = {},
  year          = {2026},
  eprint        = {2606.21519},
  archivePrefix = {arXiv},
  primaryClass  = {astro-ph.CO}
}

@book{hollander2013nonparametric,
  author        = {Hollander, M. and Wolfe, D. A. and Chicken, E.},
  title         = {{Nonparametric statistical methods}},
  edition       = {3rd},
  publisher     = {John Wiley \& Sons},
  year          = {2013}
}

@article{Ishiyama2010,
  author        = {Ishiyama, T. and Makino, J. and Ebisuzaki, T.},
  title         = {{Gamma-ray signal from Earth-mass dark matter microhalos}},
  journal       = {Astrophys. J. Lett.},
  volume        = {723},
  pages         = {L195},
  year          = {2010},
  doi           = {10.1088/2041-8205/723/2/L195},
  eprint        = {1006.3392},
  archivePrefix = {arXiv},
  primaryClass  = {astro-ph.CO}
}

@article{Anderhalden2013,
  author        = {Anderhalden, D. and Diemand, J.},
  title         = {{Density profiles of CDM microhalos and their implications for annihilation boost factors}},
  journal       = {J. Cosmol. Astropart. Phys.},
  volume        = {2013},
  number        = {04},
  pages         = {009},
  year          = {2013},
  doi           = {10.1088/1475-7516/2013/04/009},
  eprint        = {1302.0003},
  archivePrefix = {arXiv},
  primaryClass  = {astro-ph.CO}
}

@article{Ishiyama2014,
  author        = {Ishiyama, T.},
  title         = {{Hierarchical formation of dark matter halos and the free streaming scale}},
  journal       = {Astrophys. J.},
  volume        = {788},
  pages         = {27},
  year          = {2014},
  doi           = {10.1088/0004-637X/788/1/27},
  eprint        = {1404.1650},
  archivePrefix = {arXiv},
  primaryClass  = {astro-ph.CO}
}

@article{Ogiya2016,
  author        = {Ogiya, G. and Nagai, D. and Ishiyama, T.},
  title         = {{Dynamical evolution of primordial dark matter haloes through mergers}},
  journal       = {Mon. Not. R. Astron. Soc.},
  volume        = {461},
  pages         = {3385},
  year          = {2016},
  doi           = {10.1093/mnras/stw1551},
  eprint        = {1604.02866},
  archivePrefix = {arXiv},
  primaryClass  = {astro-ph.CO}
}

@article{Angulo2017,
  author        = {Angulo, R. E. and Hahn, O. and Ludlow, A. D. and Bonoli, S.},
  title         = {{Earth-mass haloes and the emergence of NFW density profiles}},
  journal       = {Mon. Not. R. Astron. Soc.},
  volume        = {471},
  pages         = {4687},
  year          = {2017},
  doi           = {10.1093/mnras/stx1658},
  eprint        = {1604.03131},
  archivePrefix = {arXiv},
  primaryClass  = {astro-ph.CO}
}

@article{OgiyaHahn2018,
  author        = {Ogiya, G. and Hahn, O.},
  title         = {{What sets the central structure of dark matter haloes?}},
  journal       = {Mon. Not. R. Astron. Soc.},
  volume        = {473},
  pages         = {4339},
  year          = {2018},
  doi           = {10.1093/mnras/stx2639},
  eprint        = {1707.07693},
  archivePrefix = {arXiv},
  primaryClass  = {astro-ph.CO}
}

@article{IshiyamaAndo2020,
  author        = {Ishiyama, T. and Ando, S.},
  title         = {{The abundance and structure of subhaloes near the free streaming scale and their impact on indirect dark matter searches}},
  journal       = {Mon. Not. R. Astron. Soc.},
  volume        = {492},
  pages         = {3662},
  year          = {2020},
  doi           = {10.1093/mnras/staa069},
  eprint        = {1907.03642},
  archivePrefix = {arXiv},
  primaryClass  = {astro-ph.CO}
}

@article{Colombi2021,
  author        = {Colombi, S.},
  title         = {{Phase-space structure of protohalos: Vlasov versus particle-mesh}},
  journal       = {Astron. Astrophys.},
  volume        = {647},
  pages         = {A66},
  year          = {2021},
  doi           = {10.1051/0004-6361/202039719},
  eprint        = {2012.04409},
  archivePrefix = {arXiv},
  primaryClass  = {astro-ph.CO}
}

@article{Tambalo2023,
  author        = {Tambalo, G. and Zumalac{\'a}rregui, M. and Dai, L. and Cheung, M. H.-Y.},
  title         = {{Gravitational wave lensing as a probe of halo properties and dark matter}},
  journal       = {Phys. Rev. D},
  volume        = {108},
  pages         = {103529},
  year          = {2023},
  doi           = {10.1103/PhysRevD.108.103529},
  eprint        = {2212.11960},
  archivePrefix = {arXiv},
  primaryClass  = {astro-ph.CO}
}

@article{Singh2025,
  author        = {Singh, S. and Brando, G. and Savastano, S. and Zumalac{\'a}rregui, M.},
  title         = {{Gravitational wave lensing: probing fuzzy dark matter with LISA}},
  journal       = {J. Cosmol. Astropart. Phys.},
  volume        = {2025},
  number        = {07},
  pages         = {025},
  year          = {2025},
  doi           = {10.1088/1475-7516/2025/07/025},
  eprint        = {2502.10758},
  archivePrefix = {arXiv},
  primaryClass  = {astro-ph.CO}
}

@article{Urrutia2025,
  author        = {Urrutia, J. and Vaskonen, V.},
  title         = {{Dark timbre of gravitational waves}},
  journal       = {Phys. Rev. D},
  volume        = {111},
  pages         = {123047},
  year          = {2025},
  doi           = {10.1103/PhysRevD.111.123047},
  eprint        = {2402.16849},
  archivePrefix = {arXiv},
  primaryClass  = {gr-qc}
}

@article{Zumalacarregui2024,
  author        = {Zumalac{\'a}rregui, M.},
  title         = {{Lens stochastic diffraction: a signature of compact objects in gravitational-wave data}},
  journal       = {},
  year          = {2024},
  eprint        = {2404.17405},
  archivePrefix = {arXiv},
  primaryClass  = {gr-qc}
}

@article{Choi2026,
  author        = {Choi, H. G. and Urrutia, J. and Zumalac{\'a}rregui, M.},
  title         = {{Signatures of $10-10^4\,{\rm M}_{\odot}$ dark matter halos in LISA via stochastic diffraction}},
  journal       = {},
  year          = {2026},
  eprint        = {2607.11887},
  archivePrefix = {arXiv},
  primaryClass  = {astro-ph.CO}
}

@article{wilcoxon1945individual,
  author        = {Wilcoxon, F.},
  title         = {{Individual comparisons by ranking methods}},
  journal       = {Biometrics Bull.},
  volume        = {1},
  pages         = {80},
  year          = {1945}
}
\appendix
\end{document}